\documentclass[pdf,twocolumn,epjc3]{svjour3}
\RequirePackage[T1]{fontenc}
\usepackage[utf8]{inputenc}
\usepackage{amsmath}
\usepackage{cite}
\usepackage{hyperref}
\usepackage[pdftex]{graphicx}
\usepackage[colorinlistoftodos,prependcaption,textsize=tiny]{todonotes}
\usepackage{subfigure}
\usepackage{amsmath}
\usepackage{color}
\usepackage{xspace}

\newcommand{\GeV}{\ensuremath{\,\mathrm{GeV}}\xspace}
\newcommand{\TeV}{\ensuremath{\,\mathrm{TeV}}\xspace}

\newcommand{\bea}{\begin{eqnarray}}
\newcommand{\eea}{\end{eqnarray}}
\newcommand{\HjjnloPS}{\ensuremath{h(2^\star)\bigoplus {\rm PS}}~}
\newcommand{\HjjjnloPS}{\ensuremath{h(3^\star)\bigoplus {\rm PS}}~}

\journalname{Eur. Phys. J. C}

\begin{document}


\title{NLO Multijet Merging for Higgs Production Beyond the VBF Approximation}

\author{Tinghua Chen\thanksref{e1,addr1}
\and Terrance M. Figy\thanksref{e2,addr1}
\and Simon Pl\"atzer\thanksref{e3,addr2,addr3,addr4}}

\thankstext{e1}{e-mail: txchen2@shockers.wichita.edu }
\thankstext{e2}{e-mail: terrance.figy@wichita.edu}
\thankstext{e3}{e-mail: simon.plaetzer@uni-graz.at}
\institute{Department of Mathematics, Statistics and Physics, Wichita State University, 1845 Fairmount St, Wichita, Kansas 67260, USA \label{addr1} \and  Institute of Physics, NAWI Graz, University of Graz, Universit\"atsplatz 5, A-8010 Graz, Austria\label{addr2} \and
  Particle Physics, Faculty of Physics, University of Vienna,
  Boltzmanngasse 5, A-1090 Wien, Austria \label{addr3}\and Erwin Schr\"odinger Institute for Mathematics and Physics, University of Vienna, Boltzmanngasse 9, A-1090 Wien, Austria\label{addr4}}
\date{Received: date / Accepted: date}
\maketitle

\abstract{We present results of the simulation of electroweak Higgs boson production at the Large Hadron Collider using both the NLO multijet merging and NLO matching frameworks provided by the general purpose event generator {\tt Herwig 7}. For the simulation of the hard scattering processes, we use the {\tt HJets} library for the full calculation and {\tt VBFNLO} for the approximate calculation to compute the $2\to h+n$ amplitudes at tree-level with $n=2,3,4$ and at one-loop with $n=2,3$. 
}

\section{Introduction} 
\label{sec:Introduction}

\begin{figure}
    \centering
    \subfigure[]{
    \label{fig:tchannel}
    \includegraphics[scale=0.2]{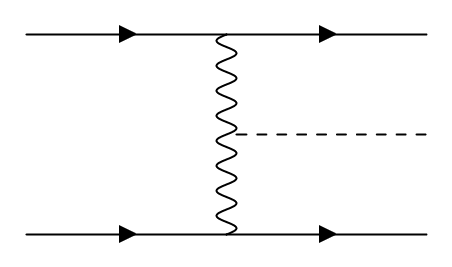}
    \includegraphics[scale=0.2]{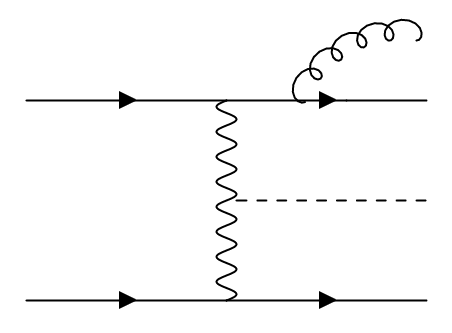}
    }
    \hfill
    \subfigure[]{
    \includegraphics[scale=0.2]{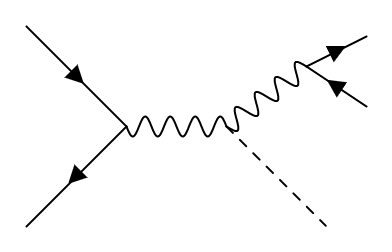}
    \includegraphics[scale=0.2]{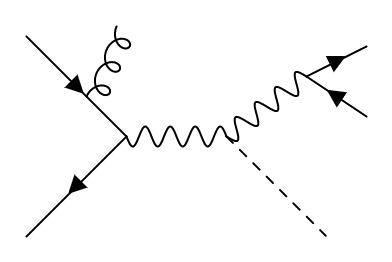}
    \label{fig:schannel}
    }
    \caption{Feynman diagrams for $t$-channel (top) and $s$-channel (bottom), where solid arrowed lines represent quarks, dotted line denotes Higgs boson, wavy lines are electroweak bosons, and curly lines are gluons.}
    \label{fig:tree}
\end{figure}

The Standard Model~\cite{Weinberg:1967tq,Salam:1968rm,Glashow:1961tr} provides a theoretical basis for understanding the fundamental origin of both fermion and electroweak gauge boson masses via the Higgs mechanism~\cite{Kibble:1967sv,Guralnik:1964eu,Englert:1964et,Higgs:1964ia,Higgs:1964pj,Higgs:1966ev}. 
The Vector-Boson Fusion (VBF) production channel that results from the coupling of a single Higgs boson to gauge bosons is one of the most important Higgs boson production channels since it serves as a probe of the mechanism for spontaneous symmetry breaking~\cite{Kauer:2000hi,Rainwater:1997dg,Rainwater:1998kj,Rainwater:1999sd,Asai:2004ws,Cranmer:2004uz,Eboli:2000ze}. 
The underlying hard scattering processes contributing to the production of a Higgs boson at $\mathcal{O}(\alpha^3)$ accompanied by at least two jets in the final state consist of space-like, $t$-channel (VBF), weak gauge boson exchange (see Fig.~\ref{fig:tchannel}) in addition to time-like, $s$-channel (VH), weak gauge boson exchange (see Fig.~\ref{fig:schannel}) topologies. 
The region of phase space where the $t$-channel topologies dominate is referred to as the VBF region. The VBF region is defined by the requirement of two energetic jets, well separated in rapidity and with the Higgs boson decay products located in the central region of the detector and possibly in the rapidity region between the two jets. In order to further reduce backgrounds, a veto on additional QCD activity occurring in the central rapidity region between the two jets can be applied to enrich the contribution of the colour singlet $t$-channel topologies~\cite{Eboli:2000ze,Cranmer:2004uz,Asai:2004ws,Rainwater:1999sd,Rainwater:1997dg,Rainwater:1998kj,Kauer:2000hi}.

\begin{figure}
    \centering
    \subfigure[]{\includegraphics[scale=0.2]{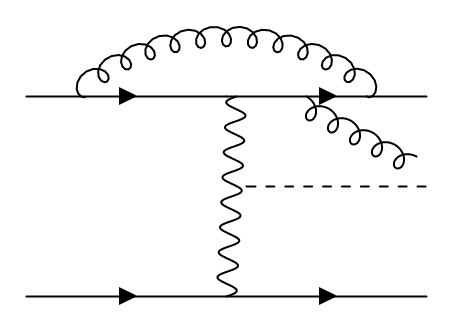}}
    \subfigure[]{\includegraphics[scale=0.2]{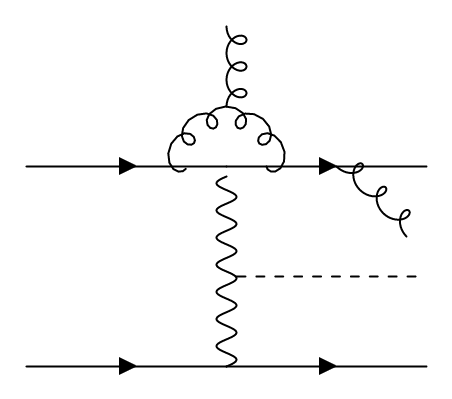}}
    \subfigure[]{\includegraphics[scale=0.2]{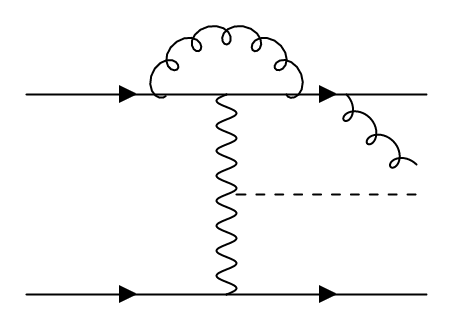}}
    \subfigure[]{\includegraphics[scale=0.2]{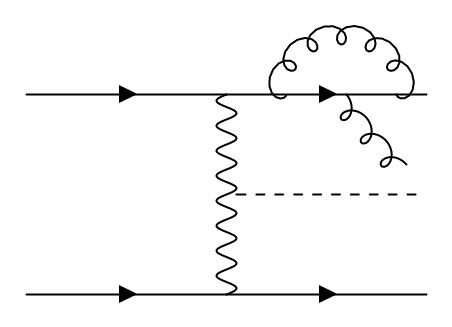}}
    \caption{Representative one-loop Feynman diagram topologies for $h+3$ jet production where the gluon connects the same quark line.}
    \label{fig:boxline}
\end{figure}

\begin{figure}
    \centering
    \subfigure[]{\includegraphics[scale=0.2]{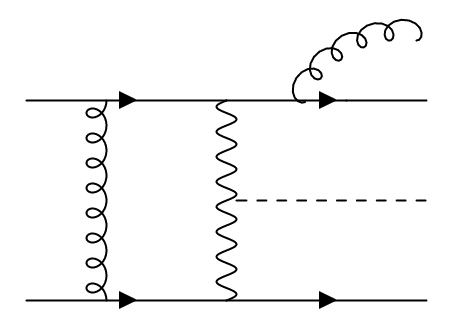}}
    \subfigure[]{\includegraphics[scale=0.2]{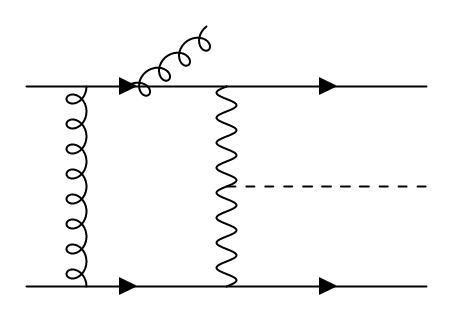}}
    \caption{Representative hexagon and pentagon one-loop Feynman diagram topologies for $h+3$ jet production.}
    \label{fig:pent}
\end{figure}

Several theoretical predictions for VBF Higgs boson production have employed the VBF approximation where only $t$-channel topologies are retained while the $s$-channel topologies are discarded in addition to interference contributions. Essentially, the VBF approximation amounts to treating the incoming hard scattering quarks as belonging to two separate $SU(3)$ colour gauge groups. At one-loop this implies that all topologies where the gluon connects two different quark lines are neglected\footnote{For the case of non-identical quarks the interference of pentagon diagrams with the Born diagrams vanishes exactly upon summing over colors. See Ref.~\cite{Ciccolini:2007ec} for more details.}. Perturbative QCD corrections for integrated scattering cross-sections for Higgs boson production were first computed in the structure function approach at NLO \cite{Han:1992hr} and later at NNLO \cite{Bolzoni:2010xr}. The perturbative QCD corrections for VBF Higgs production have been computed to ${\rm N}^{3}$LO accuracy in Ref. \cite{Dreyer:2016oyx} within the structure-function approach. Kinematic distributions in the form of a parton-level Monte Carlo program for VBF Higgs production were presented in Refs.~\cite{Figy:2003nv,Berger:2004pca}. The NLO QCD and electroweak corrections have been computed to NLO accuracy for the full set of Feynman diagrams in Ref. \cite{Ciccolini:2007ec}. The first NNLO parton-level program which uses the projection-to-Born technique for VBF Higgs production was developed by the {\tt proVBFH} collaboration in Ref.~\cite{Cacciari:2015jma} in the $t$-channel approximation.  Recently, the {\tt NNLOJet} collaboration using antenna subtraction developed a NNLO parton-level Monte Carlo program that computes the NNLO corrections for VBF Higgs production in the $t$-channel approximation~\cite{Cruz-Martinez:2018dvl,Cruz-Martinez:2018rod}.
The first calculation of Higgs boson production in association with three jets ($h+3$ jet) at NLO in the $t$-channel (VBF$+$jet) approximation was presented in Ref.~\cite{Figy:2007kv} using one-loop Feynman diagrams depicted in Fig.~\ref{fig:boxline} and neglecting one-loop pentagon and hexagon Feynman diagrams depicted in Fig.~\ref{fig:pent}.
VBF Higgs production has been simulated at NLO with a parton shower by several groups (see Refs. \cite{Jager:2020hkz,Nason:2009ai,Frixione:2013mta,DErrico:2011wfa,Azzi:2019yne}). 
VBF$+$jet production at NLO matched to a parton shower has been investigated in Refs.~\cite{Jager:2014vna,deFlorian:2016spz}. 
Recently, after applying strict VBF selection cuts, differential cross sections have been evaluated for VBF Higgs production using high multiplicity partonic final state matrix elements provided by {\tt Sherpa MC}\cite{Sherpa:2019gpd} and {\tt POWHEG Box}\cite{Alioli:2010xd} within the merging framework of the {\tt Vincia} parton shower program \cite{Hoche:2021mkv}. 

The first full calculation which includes hexagon and pentagon Feynman diagrams (see Fig.~\ref{fig:pent}) in addition to the one-loop diagrams depicted in Fig.~\ref{fig:boxline} and all possible interferences between $s$-channel and $t/u$-channel type Feynman diagrams at one-loop for $h+3$ jet production at NLO was presented in Ref.~\cite{Campanario:2013fsa}.  
Subsequently, the first comparative analysis of the full $h+3$ jet production at NLO against VBF $h+3$ jet at NLO was presented in Ref.~\cite{Campanario:2018ppz}. Recently, the impact of non-factorizable corrections has been investigated in the eikonal approximation by several groups \cite{Liu:2019tuy,Dreyer:2020urf}.

Recent experimental analyses employ relaxed selection criteria for the VBF region such as defined by the STXS framework~\cite{Berger:2019wnu} and rely on a multitude of multi-variate analysis techniques instead of standard cut based analyses (see for example Ref.~\cite{ATLAS:2020bhl}). In a recent comparative study predictions for VBF Higgs boson production using the well-known VBF selection criterion \footnote{Typical VBF selection criterion requires two well separate jets with large dijet invariant mass. An example is the TIGHT selection cuts discussed in Section~\ref{sec:cuts}.} were generated using different general-purpose event generators each employing a differing choices for the NLO parton shower matching scheme and parton shower~\cite{Jager:2020hkz}. While predictions for two jet observables were NLO accurate, predictions for three jet observables were only LO accurate~\cite{Jager:2020hkz}. 
In Ref.~\cite{Campanario:2018ppz}, the authors showed for the fixed order NLO predictions of $h+3$ jet production that a significant part of the QCD corrections can be attributed to Higgs-Strahlung-type topologies. 
While predictions using the VBF selection criterion are currently very relevant, it is of theoretical interest to study the impact of relaxing the VBF selection criterion. An example is the recent comprehensive study that explored Higgs boson production in the high transverse momentum region of the Higgs boson~\cite{Buckley:2021gfw}. Hence, the focus of this paper is to use the matrix elements provided by the complete NLO $h+2$ and $h+3$ jet production calculations to predict Higgs boson and jet observables using the NLO multijet merging framework provided by the general purpose event generator {\tt Herwig 7}. 

This paper is organized as follows. In Section~\ref{sec:setup} we discuss the settings and input parameters that we used for simulating the production of a Higgs boson using the {\tt Herwig 7} general purpose Monte Carlo program and we discuss the matrix elements used to model the hard scattering processes. In Section~\ref{sec:results} we present predictions for several relevant Higgs and jet observables. We discuss our conclusions in Section~\ref{sec:Conclusion}.   

\section{Setup for Simulation}
\label{sec:setup}
In this section we provide a summary of the software tools, input parameters, and analysis cuts we used in our simulations. We use 
{\tt Herwig 7}~\cite{Bahr:2008pv,Bellm:2015jjp,Platzer:2011bc,Bellm:2019zci}, a multi-purpose event generator, to simulate the scattering of protons at the LHC experiment.  In this paper, we do not consider the phenomenology of Higgs boson decays and treat the Higgs boson as a stable particle in our simulations. Our simulations are performed at the parton-level, i.e., we do not include hadronization or multiple parton interactions (MPI) in our simulations.  

\subsection{Notation}
In this section we will define the notation we use in the following when discussing the results of simulations involving the merging and matching of different partonic multiplicities of tree-level and one-loop matrix elements. 

For electroweak Higgs boson production, the Born-level hard process is $pp \to hjj$, where the tree-level $2 \to h+2$ partonic matrix elements (MEs) are represented by the symbol $h(2)$. For a leading order merged setup, we add tree-level MEs for additional partonic multiplicity, such that we write $h(2,3,4, \ldots, n)$ with $n \geq 2$ for the merging of MEs with $h+2,h+3,h+4,\ldots, h+n$ partons.

For simulations involving the merging of higher-order MEs, we denote each multiplicity $n$ with an extra $\star$ as $n^{\star}$ for which, in addition to tree-level MEs, the one-loop correction is included. For example, $h(2^{\star},3,4)$ denotes Higgs boson production with up to 2 additional partonic emissions where one-loop MEs have been utilized for the $h+2$ parton processes. The special case of $h(2^{\star},3)$ describes the "matching through merging" limit of the merging approach that {\tt Herwig 7} uses which delivers the same accuracy as NLO matching.  For a discussion of merging see Subsection~\ref{sub:merging}. For scenarios where we use subtractive matching to match an NLO calculation to a dipole parton shower (see Subsection~\ref{sub:matching} for further details), we denote the NLO matched setup for $pp\to hjj$ by \\
$h(2^\star)\bigoplus {\rm PS}$ and $pp\to hjjj$ by $h(3^\star)\bigoplus {\rm PS}$.  

\subsection{Choice of Matrix Elements}
The matrix elements used in this paper were provided via the external matrix element providers: {\tt HJets 1.2} and {\tt VBFNLO 3}. The {\tt HJets 1.2} module  ~\cite{Campanario:2013fsa,Campanario:2013nca,Campanario:2014aia,Campanario:2018ppz} provides tree-level matrix elements for $2 \to h+2,3,4$ partons and one-loop matrix elements for $2 \to h+2,3$ partons for electroweak Higgs boson production in association with jets. The colour algebra is computed by the library {\tt ColorFull}~\cite{Sjodahl:2014opa} and the one-loop integrals are performed based on the tensor loop integral reduction methods described in Ref.~\cite{Campanario:2011cs}. We used the matrix elements encoded in {\tt VBFNLO 3.0.0} beta $5$~\cite{baglio2011vbfnlo,baglio2014release,Arnold:2008rz,Figy:2007kv} to compute tree-level and one-loop matrix elements in the VBF approximation via the Binoth one-loop accord~\cite{Alioli:2013nda,Binoth:2010xt}. 

\subsection{Input Parameters} 

We compute electroweak parameters in the $G_{\mu}$ scheme with the following input parameters: $ G_F=1.16637\times10^{-5}~\GeV^{-2}$, $m_Z=91.1876~\GeV$, and $m_W=80.385~\GeV$. The widths of the $Z^{0}$ and $W^{\pm}$ gauge bosons are fixed to $\Gamma_Z=2.4952~\GeV$ and $\Gamma_W=2.085~\GeV$, respectively. The electromagnetic coupling constant and the weak mixing angle are calculated at tree-level. The Higgs boson mass is set to $m_H=125.7~\GeV$.
 
For renormalization and factorization scales, we choose the following as our central scale:
\begin{equation}
	\mu_0=\frac{1}{2} H_{T,{\rm jets}}= \frac{1}{2} \sum_{i \in \text{jets}} p_{T,i},
\end{equation}
 where $p_{T,i}$ is the transverse momentum of the $i$-th jet. In order to ensure infrared safety of our scale choice, we use anti-$k_{T}$ jet clustering~\cite{Cacciari:2005hq} with $R=0.4$ in the inclusive mode and choose the {\tt E-scheme} recombination method and require that each jet have $p_{T,i}>5$ GeV. The collider energy $\sqrt{s}=13~\TeV$ is used for all simulations. For all simulations we use {\tt LHAPDF6}\cite{Buckley:2014ana} with the parton distribution functions \\
{\tt PDF4LHC15\_nnlo\_100\_pdfas}~\cite{Butterworth_2016} with $\alpha_S(M_Z)=0.118$ and five active flavours.

\subsection{Event Generation Cuts and Kinematic Selection Cuts}
\label{sec:cuts}
Since the lowest order process involves jets we apply a set of hard process selection cuts before the subsequent parton shower, however we do ensure that there is enough margin in our analysis cuts. While in principle not needed for the lowest-order, two jet process, we still use a cut for efficiency reasons; other processes of the VBF/VBS kind do require generation cuts even at NLO for the lowest order. Jet reconstruction is implemented on final state partons utilizing the anti-$k_{t}$ algorithm~\cite{Cacciari:2005hq} via the {\tt fastjet} library~\cite{Cacciari:2011ma}. We use $R=0.4$ in the inclusive mode and choose the {\tt E-scheme} for the method of recombination. The minimum transverse momentum for all jets is set to $10$ GeV with the rapidity of the jet restricted to $|y_{j}|\leq 5$. For setups where the lowest final state parton multiplicity is two partons such as $h(2,3)$ or \HjjnloPS, we require at least two jets while for setups where the lowest final state parton multiplicity is $3$ partons such as \HjjjnloPS we require at least three jets. 

All simulated events are analyzed using the MC analysis toolkit {\tt Rivet 2.7.2}~\cite{Buckley_2013} and the source code for the analyses is publicly available at GitHub \cite{HJets:merging}. Using the {\tt Rivet} framework, we developed an analysis labeled {\tt MC\_H2JETS} that implements three choices of event selection criterion: inclusive cuts (INCL), tight cuts (TIGHT), and loose cuts (LOOSE).  Partons are combined into jets using the anti-$k_{t}$ algorithm~\cite{Cacciari:2005hq} with radius parameter $R=0.4$ in the inclusive mode and perform the recombination of partons into jets in the {\tt E-scheme}. All jets are required to have a transverse momentum $p_{T,j}$ and rapidity $y_{j}$ restricted by the following conditions:
\begin{equation}
p_{T,j} > 25~\GeV,\quad |y_{j}| \leq 4.5. 
\label{eq:inclusivecut}
\end{equation}
In this paper we order jets from largest to smallest in jet transverse momentum and label jets as $j_{k}$ with $k=1,2,3...$ being an index.
For INCL selection cuts we require at least two jets in the event. 
For the LOOSE selection cuts we include the following additional selection criterion
\begin{equation}
	m_{j_1j_2}>200~\GeV,\quad \Delta y_{j_1j_2}>1,
\end{equation} 
where the invariant mass of the leading two jets is defined as $m_{j_1 j_2}=\sqrt{(p_{j_1}+p_{j_2})^2}$ and the rapidity separation of the leading two jets is defined as $\Delta y_{j_1 j_2}=|y_{j_1}-y_{j_2}|$.
For the TIGHT selection cuts we include the following additional selection criterion
\begin{equation}
	m_{j_1j_2}>600~\GeV,\quad \Delta y_{j_1j_2}>4.5, \quad y_{j_1}\cdot y_{j_2}<0.
\end{equation}
For the purpose of examining jets in the rapidity gap between the leading two jets in an event, we require jets of at least $25$ GeV of transverse momentum and a rapidity $y_{\rm gap}$ such that 
$ \min(y_{j_1},y_{j_2}) < y_{\text{gap}} < \max(y_{j_1},y_{j_2})$. Jets that satisfy the above conditions are labeled as "gap jets" and ordered from largest to smallest jet transverse momentum.

In order to investigate jet production rates at different resolution scales, we have modified the {\tt Rivet} analysis {\tt MC\_HKTSPLITTINGS} to allow for a stable Higgs boson in the event simulation.  We use the $k_{T}$ algorithm \cite{Catani:1993hr,Ellis:1993tq} with $R=0.6$ to compute differential distributions of the splitting scales $\sqrt{d_{k}}$. The algorithm for the computation of jet resolution scales is discussed in detail in the introduction of Ref.~\cite{ATLAS:2013nef}. The results of this analysis are discussed in Subsection~\ref{sub:results-1}.

\subsection{The Matching Algorithm} 
\label{sub:matching}
Matching parton showers to next-to-leading order (NLO) QCD has become the de-facto standard for reliable simulations at hadron colliders. By matching we refer to a scheme which subtracts the expansion of the parton shower to ${\cal O}(\alpha_s)$ from the fixed-order NLO such that the distribution which results after showering is correct at NLO, to ${\cal O}(\alpha_s)$. 
Scale setting in the matching algorithms in Herwig 7 have been discussed in great detail in \cite{Bellm:2016rhh}, with all of these options available for both the {\tt Herwig 7} parton showers \cite{Platzer:2011bc,Platzer:2009jq,Gieseke:2003rz}. In our work we choose the default setting of the `resummation profile' which features a narrowly smeared step function towards the hard scale and thus does not introduce spuriously small variations.

At the level of the matched fixed-order cross section, {\tt Herwig 7} provides a framework for interfacing external matrix element providers such as {\tt HJets} and {\tt VBFNLO 3} to the {\tt Matchbox} module~\cite{Platzer:2011bc}, which will assemble full-fledged, differential NLO cross section from the external matrix elements. Through the {\tt Matchbox} module {\tt Herwig 7} is able to match NLO matrix elements to both the angular ordered \cite{Gieseke:2003rz} and dipole showers \cite{Platzer:2009jq,Platzer:2011bc} using either a subtractive (MC@NLO-type\cite{Frixione:2002ik}) matching algorithm or a multiplicative (Powheg-type~\cite{Nason:2004rx}) matching algorithm. In our work we focus on the dipole shower matched via the subtractive matching paradigm, since the main objective of our work is to employ merging algorithms of different jet multiplicities, which, inside of {\tt Herwig 7} are currently only available with the dipole shower algorithm. Differences between shower algorithms and matching schemes have extensively been discussed in \cite{Buckley:2021gfw} in the context of Higgs production via VBF.

\subsection{The Unitary Merging Algorithm}
\label{sub:merging}
As opposed to matching, merging algorithms do facilitate the combination of several jet multiplicities with the parton shower. At leading order, stability with respect to the resolution which separates hard jet production from parton shower radiation is achieved by carefully crafting this resolution to be compatible with the shower phase space and ordering. No spurious logarithms of the merging scale are then expected to arise, since the shower is considered to be a good approximation to tree-level real emission matrix elements in the transition region. This is not true anymore at NLO, and a new paradigm of merging needs to be employed which is correcting for the lack of perturbative information contained in the parton shower. These unitarized merging algorithms preserve certain features of inclusive cross sections \cite{Lonnblad:2011xx,Lonnblad:2012ix,Lonnblad:2012ng}, and thus generate approximate NNLO contributions which are required for a stable merging. The full implementation of the unitary merging algorithm used by {\tt Herwig 7} was described in Ref.~\cite{Bellm:2017ktr}, and does not enforce the reproduction of inclusive cross sections exactly, but only subtracts contributions which are classified as logarithmically enhanced if they are accessible by a possible parton shower history. The process is otherwise considered to contribute a new, hard jet configuration from which the parton shower is evolving in a vetoed manner such as not to double count contributions both in real emission as well as virtual and unresolved corrections. At the same time this approach allows the merging algorithm to deal with processes that involve jets already at the level of the hard process, something other merging schemes have not reported on\footnote{One exception is Ref.~\cite{Hoche:2021mkv} which appeared while this work was being finalized.}.
In our studies, the merging scale $\rho_{s}$ is smeared according to eq.(40), $\rho_s=\rho_C\cdot (1+(2\cdot r-1)\cdot \delta)$ of Ref.\cite{Bellm:2017ktr} . The central merging scale is set as $\rho_{C}=25~\GeV$ with $\delta=0.1$ for results investigating the impact of varying the factorization and renormalization scales in Subsection~\ref{sub:fr}.  We use the CMW scheme~\cite{Catani:1990rr,Bellm:2017ktr} for the merged simulations with the modified strong coupling set to $\alpha^{\prime}_{S}(q)=\alpha_{S}(k_{g}(q))$ where $k_{g}=\exp(-K_{g}/b_{0})$, $K_g=C_A\Big(\frac{67}{18}-\frac{1}{6}\pi^2\Big)-\frac{5}{9}N_F$, and  $b_0=11-2/3N_{F}$\footnote{The matched setups also employ the CMW scheme.}. 
%
\begin{figure*}
\centering 
\includegraphics[width=0.45\textwidth]{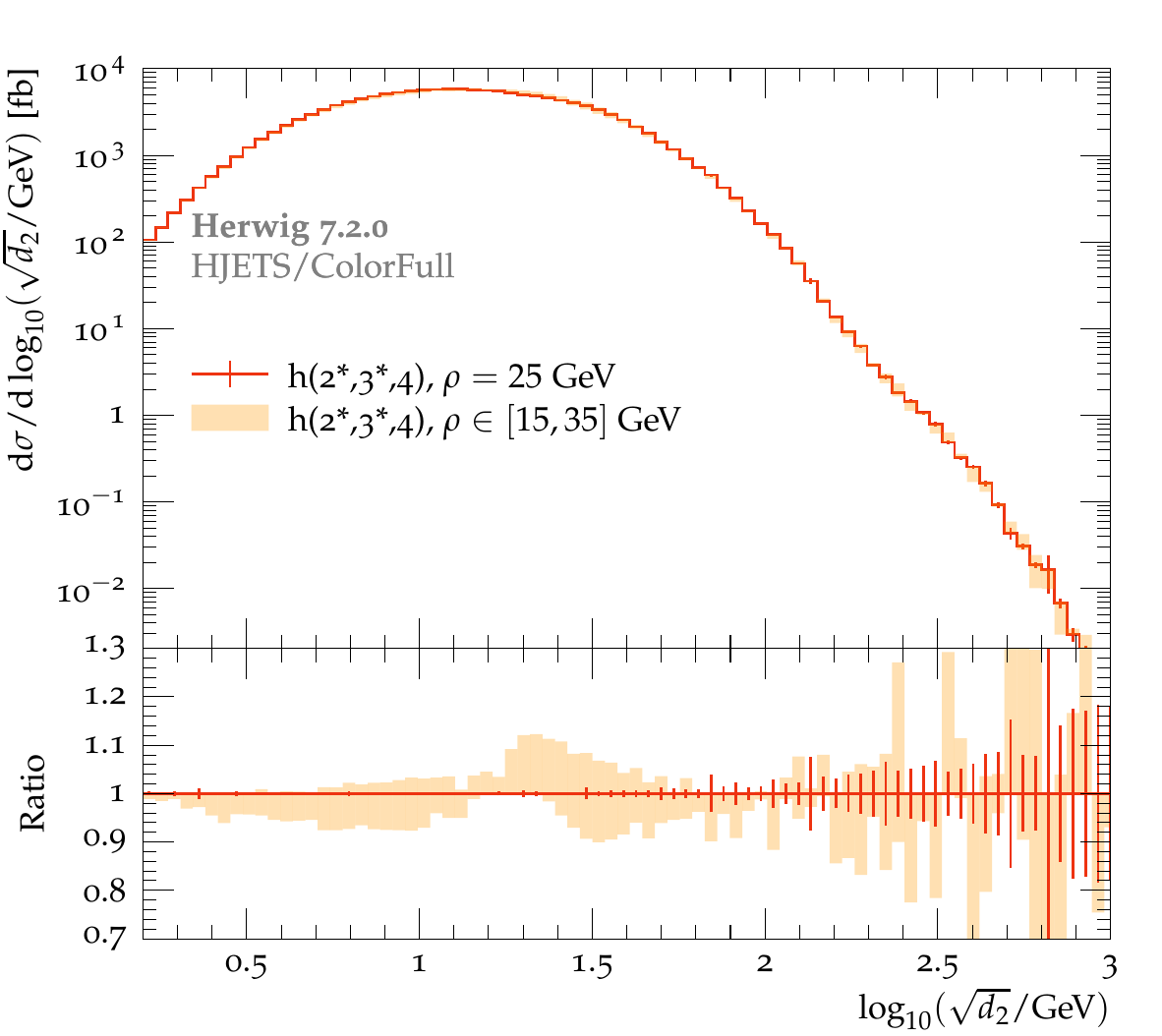}   
\includegraphics[width=0.45\textwidth]{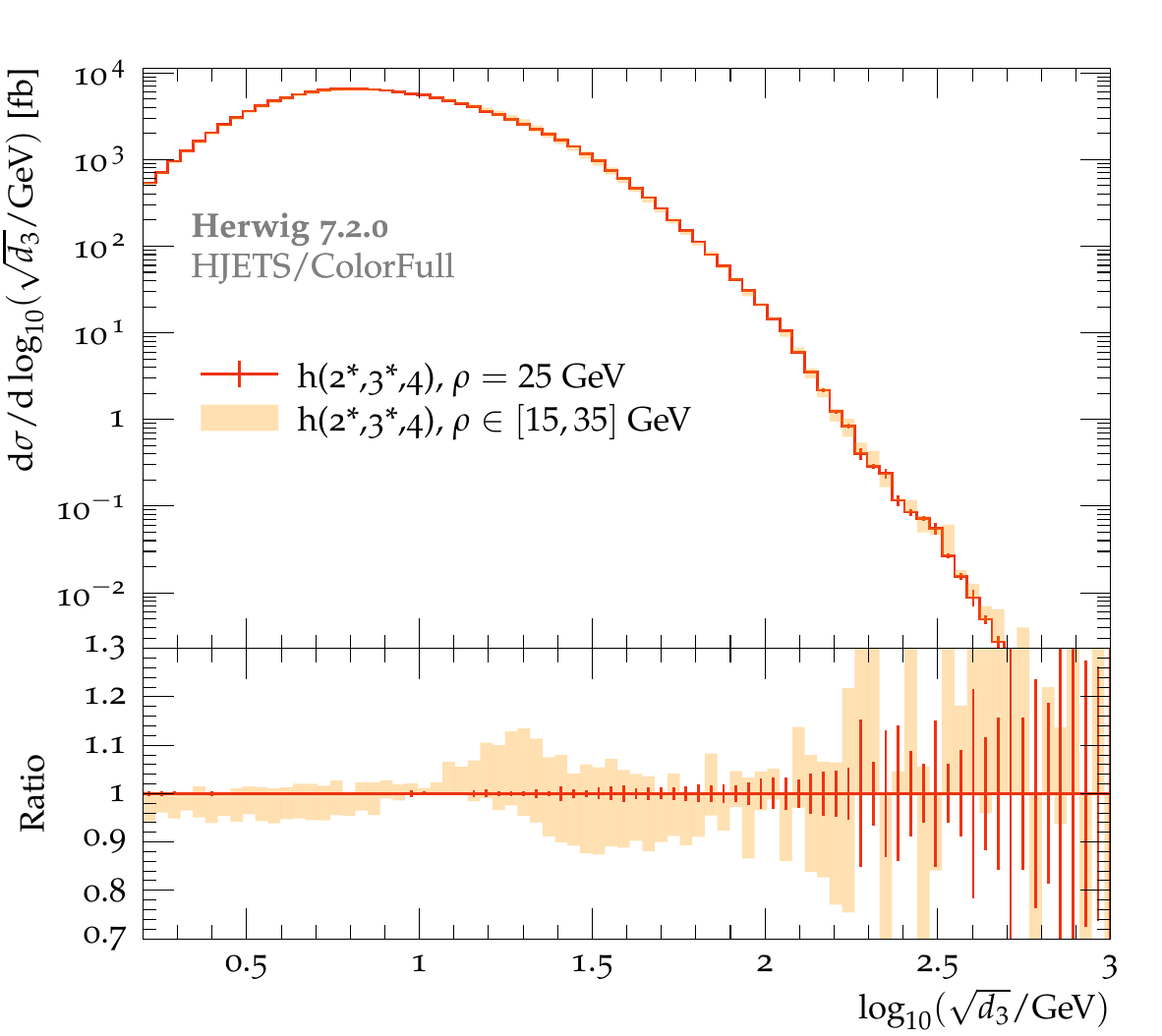}\\
\includegraphics[width=0.45\textwidth]{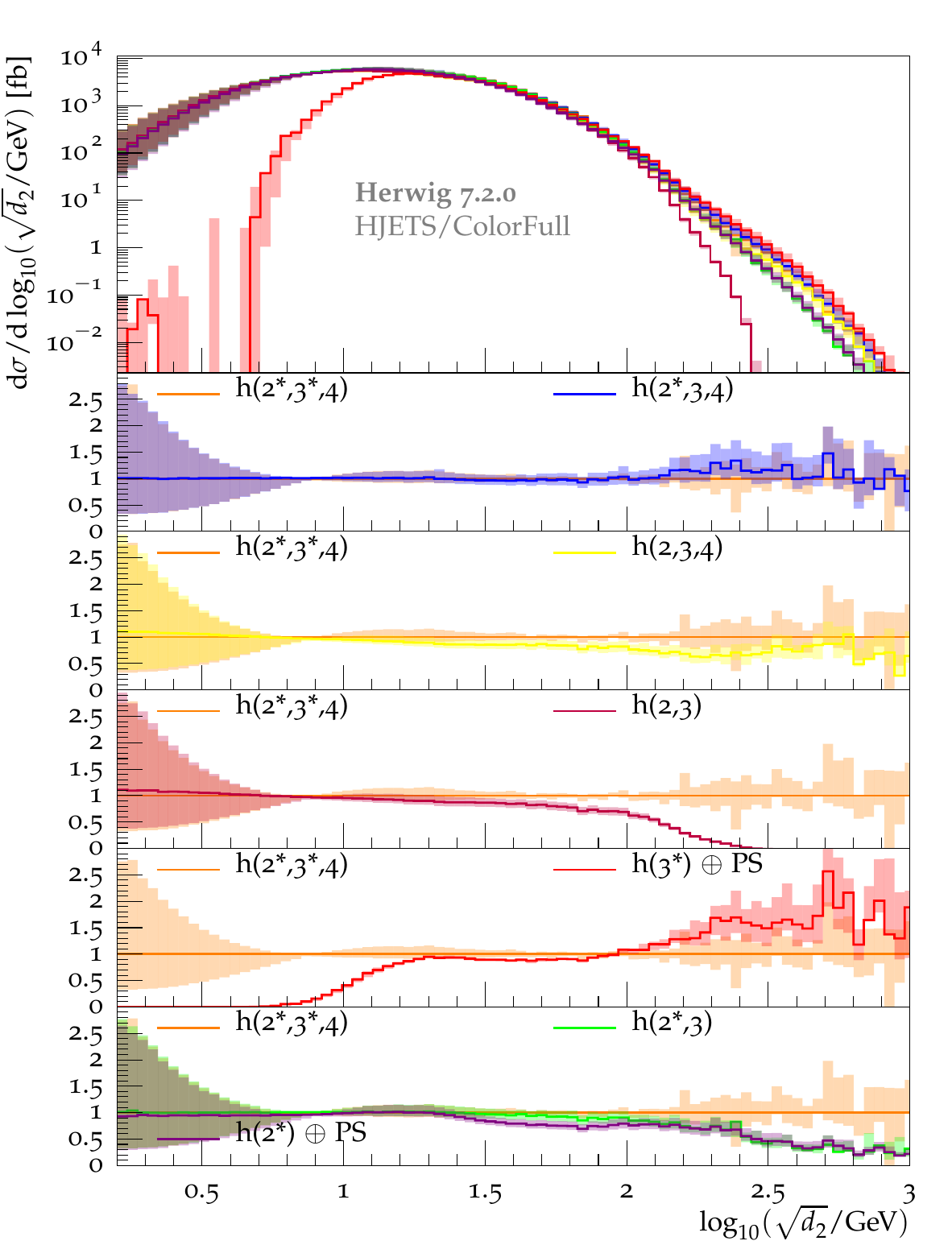}       
\includegraphics[width=0.45\textwidth]{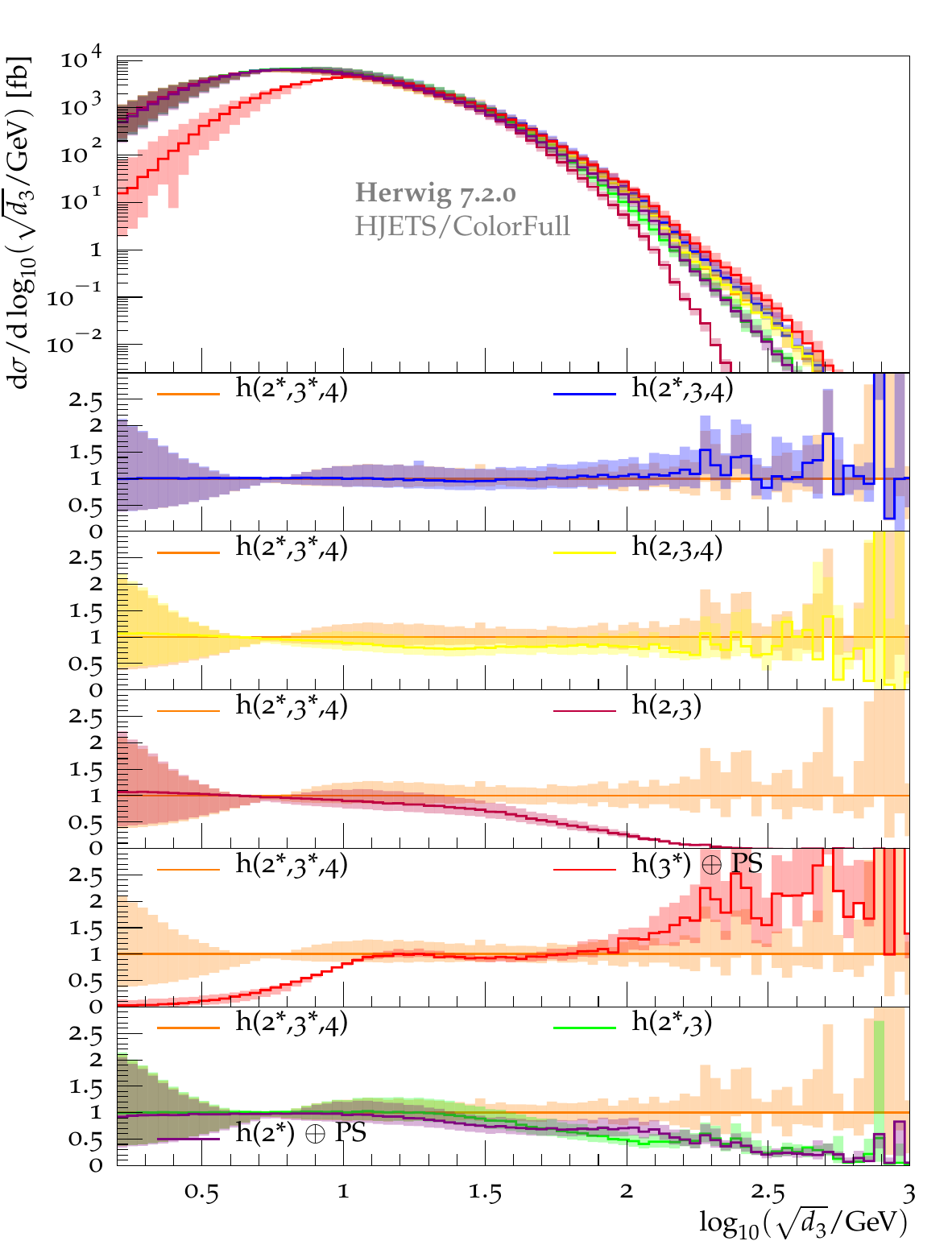} 
\caption{Comparison plots for $\sqrt{d_{2}}$ (left column) and $\sqrt{d_{3}}$ (right column). Shown in the top row is the theory band resulting from varying the merging scale for the prediction of $h(2^{\star},3^{\star},4)$. Shown in the bottom row are theory error bands resulting from the variation of the renormalization and factorization scales for predictions of $h(2^{\star},3^{\star},4)$, $h(2^{\star},3,4)$, $h(2,3,4)$, $h(2^{\star},3)$, $h(2,3)$, $h(3^{\star})\bigoplus {\rm PS}$, and $h(2^{\star}) \bigoplus {\rm PS}$ with $h(2^{\star},3^{\star},4)$ being the reference in the ratio plots. While for the observables we consider later the matched and merged three jet simulations agree well within their uncertainties we here observe bigger differences for large jet separation scales and large third jet transverse momenta. This signals that the multi-scale approach inherent to the merging is more appropriate to such processes than using a fixed-multiplicity, dynamic scale like the $H_{T,{\rm jets}}$ scale we have been using here. Also observe that additional hard jet matrix elements, like the four jet contribution, might well affect transition rates at lower jet multiplicity, a feature which has reportedly been observed in merged simulations even up to much more than a hand full of jets \cite{Hoche:2019flt} and is a consequence of the transition of four jet matrix elements into the parton shower region.}
\label{fig:ktspl}
\end{figure*}
\begin{figure*}
    \centering
        \includegraphics[width=0.45\textwidth]{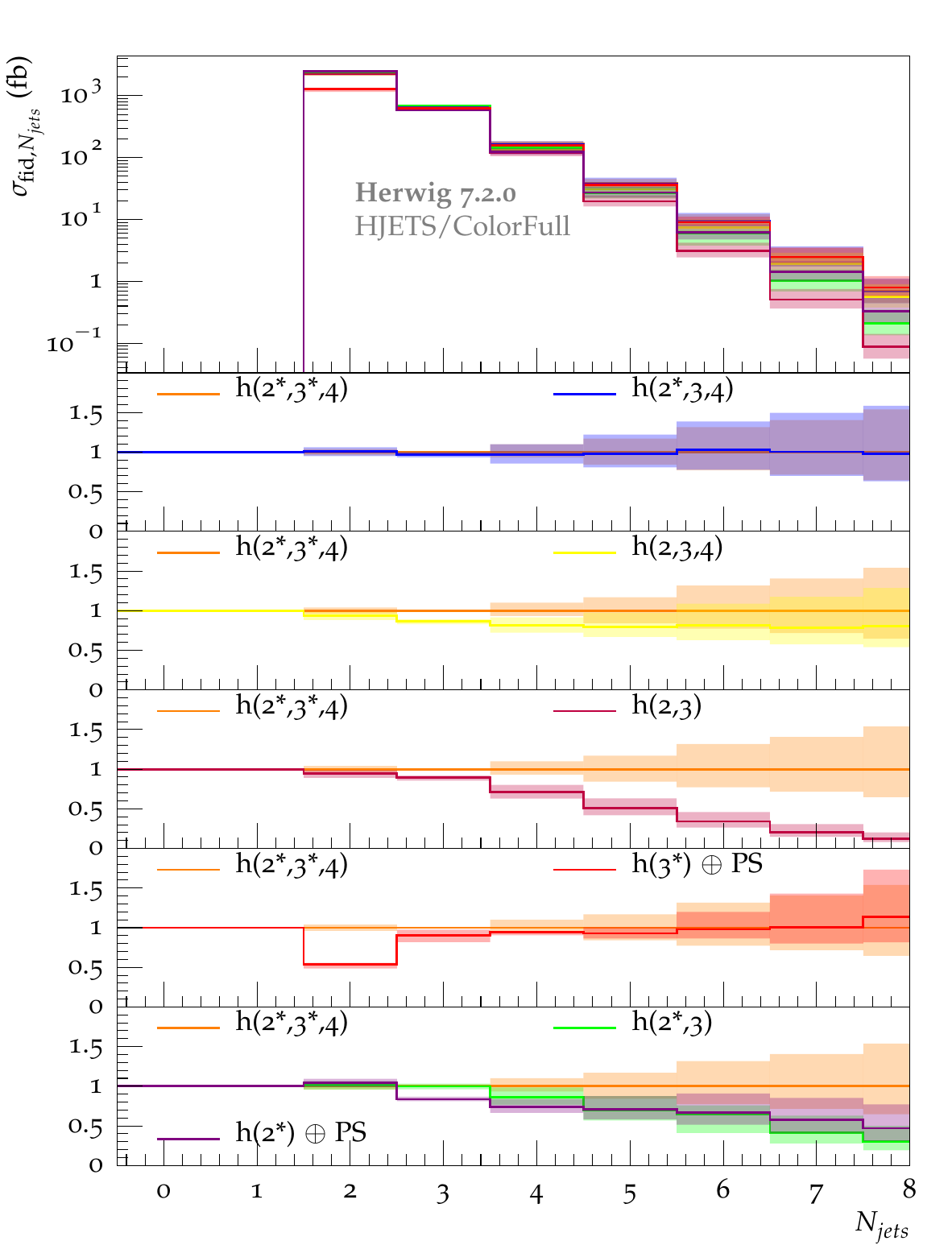} 
        \includegraphics[width=0.45\textwidth]{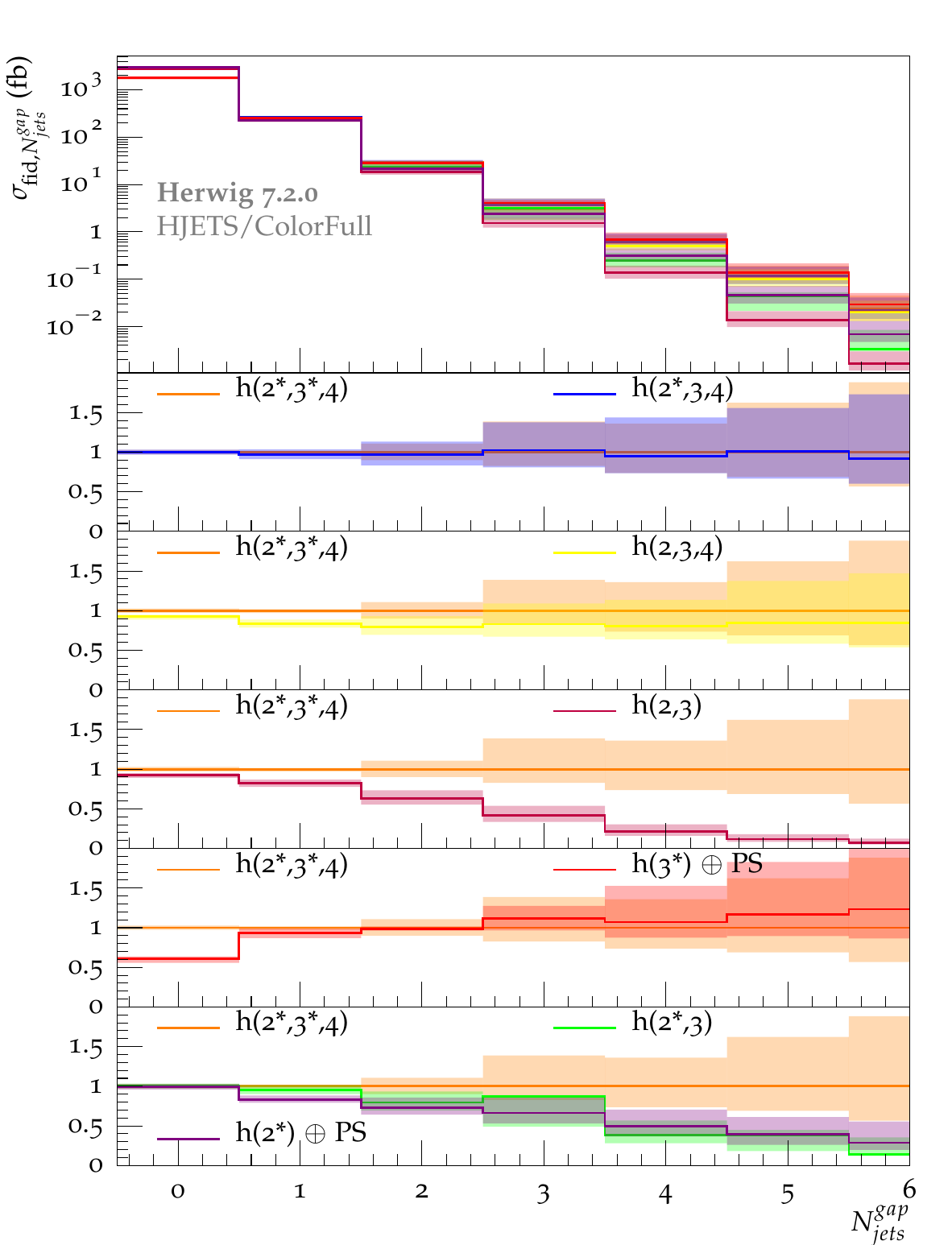} \\
        \includegraphics[width=0.45\textwidth]{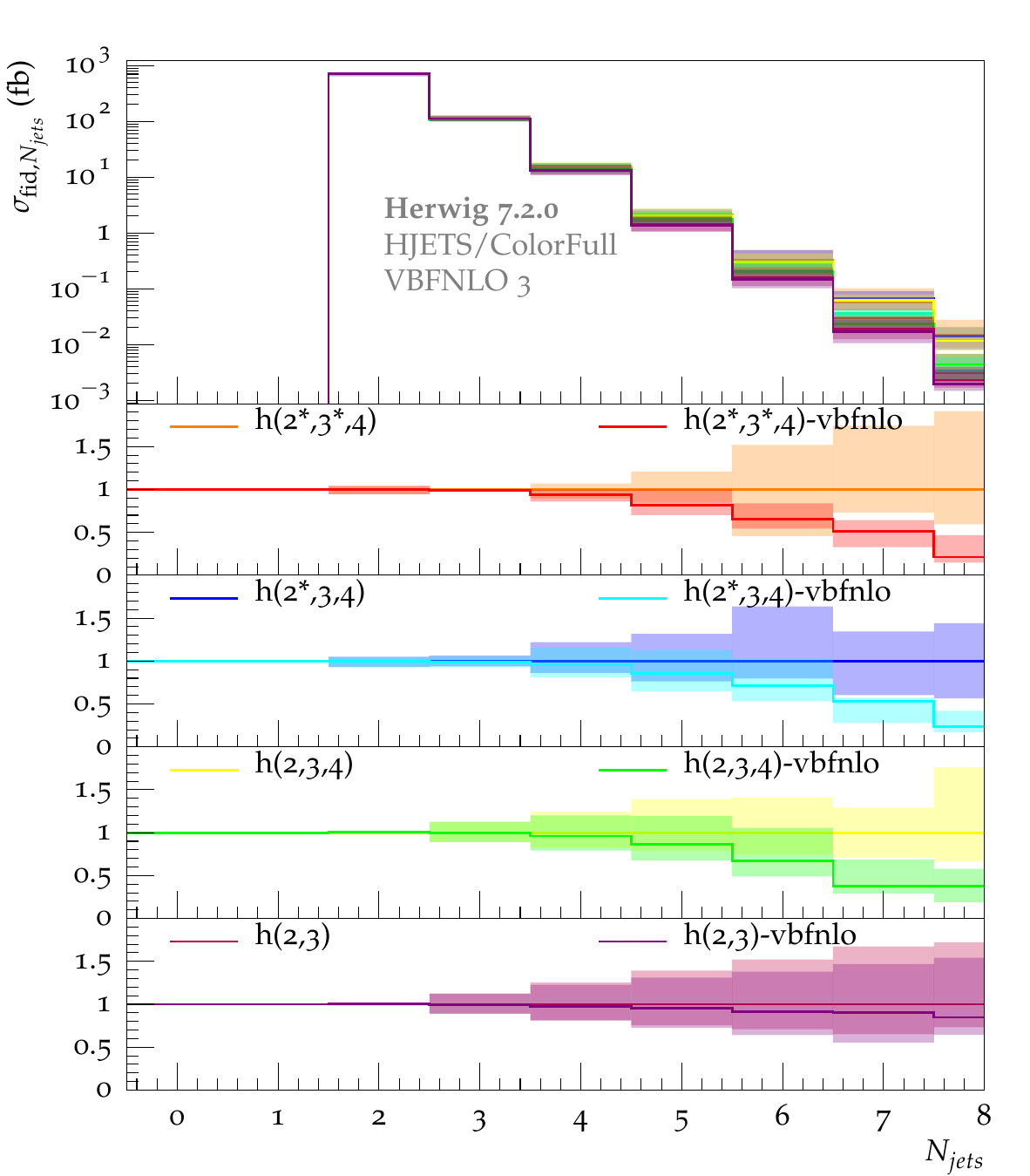} 
        \includegraphics[width=0.45\textwidth]{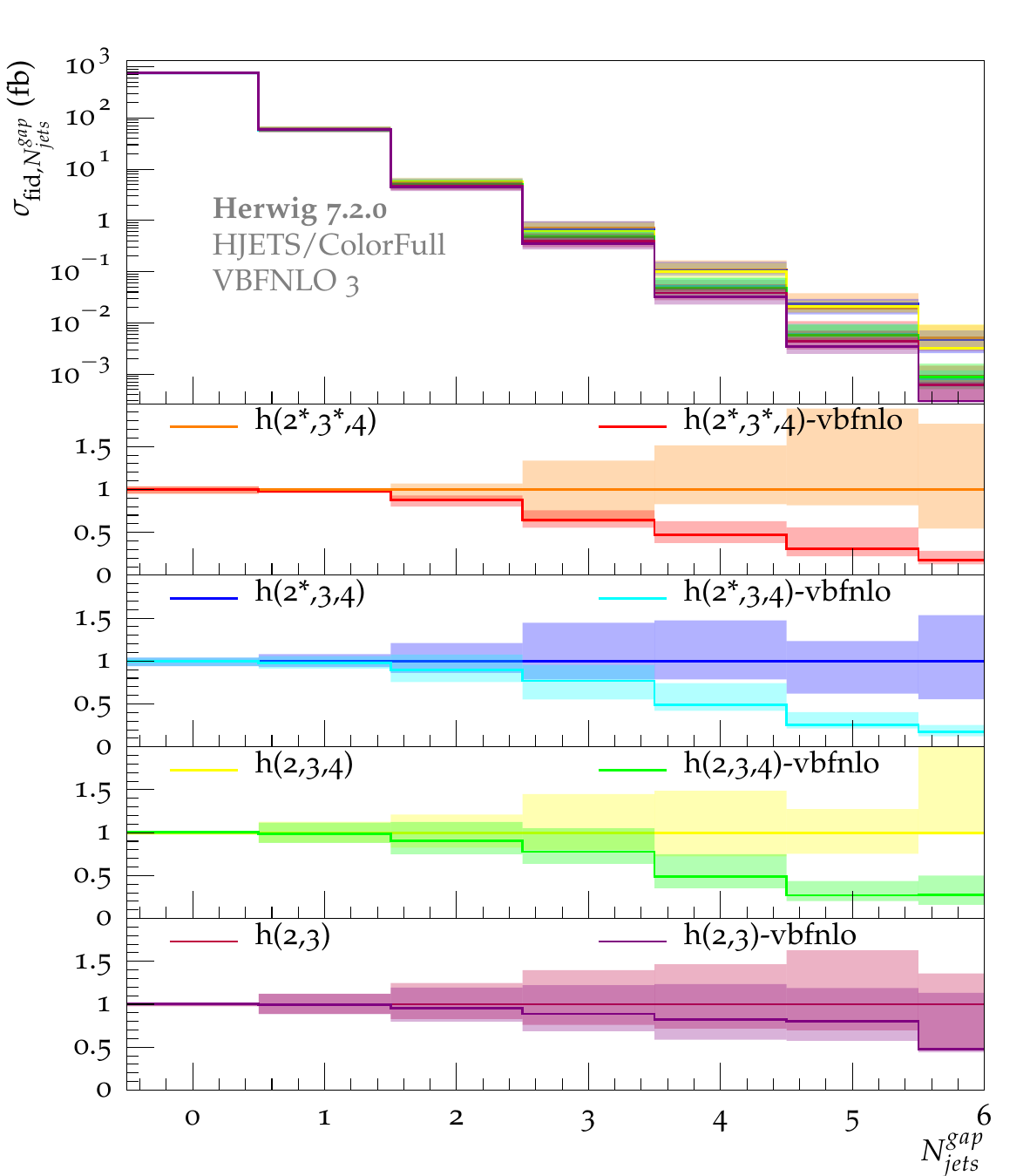}
    \caption{Exclusive number of jets (left column) and the exclusive number of gap jets (right column) with INCL (top row) and TIGHT (bottom row) selection cuts for the predictions of $h(2^{\star},3^{\star},4)$, $h(2^{\star},3,4)$, $h(2,3,4)$, $h(2^{\star},3)$, $h(2,3)$, $h(3^{\star})\bigoplus {\rm PS}$, and $h(2^{\star}) \bigoplus {\rm PS}$. $h(2^{\star},3^{\star},4)$ being the reference in the ratio plots for top row. In the bottom row setups using {\tt VBFNLO} MEs are compared against setups using {\tt HJets} MEs. Shown in the figures are theory error bands due to the variation of the factorization and renormalization scales.}
    \label{fig:xs}
\end{figure*}

\begin{figure*}
    \centering
        \includegraphics[width=0.45\textwidth]{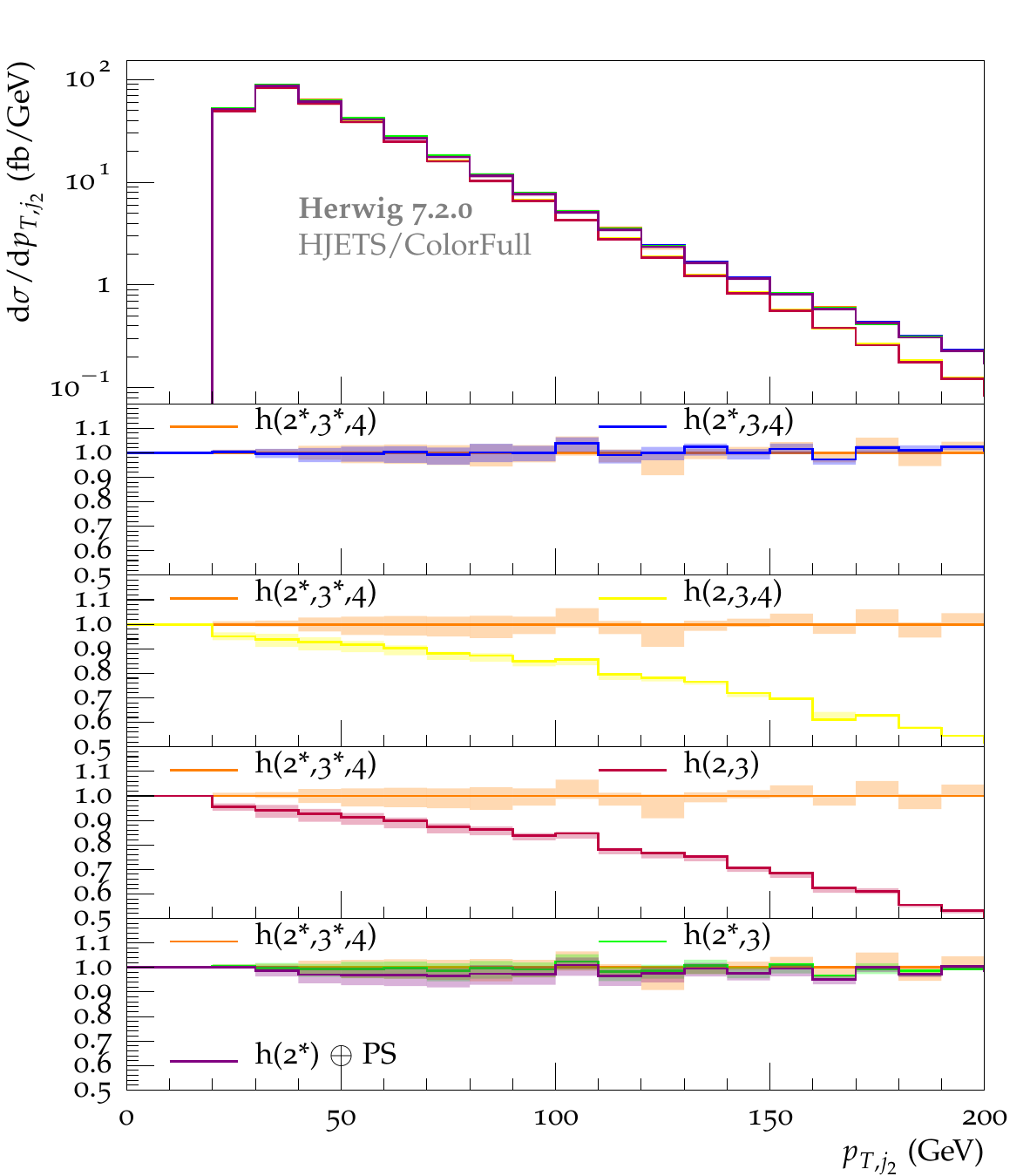} 
        \includegraphics[width=0.45\textwidth]{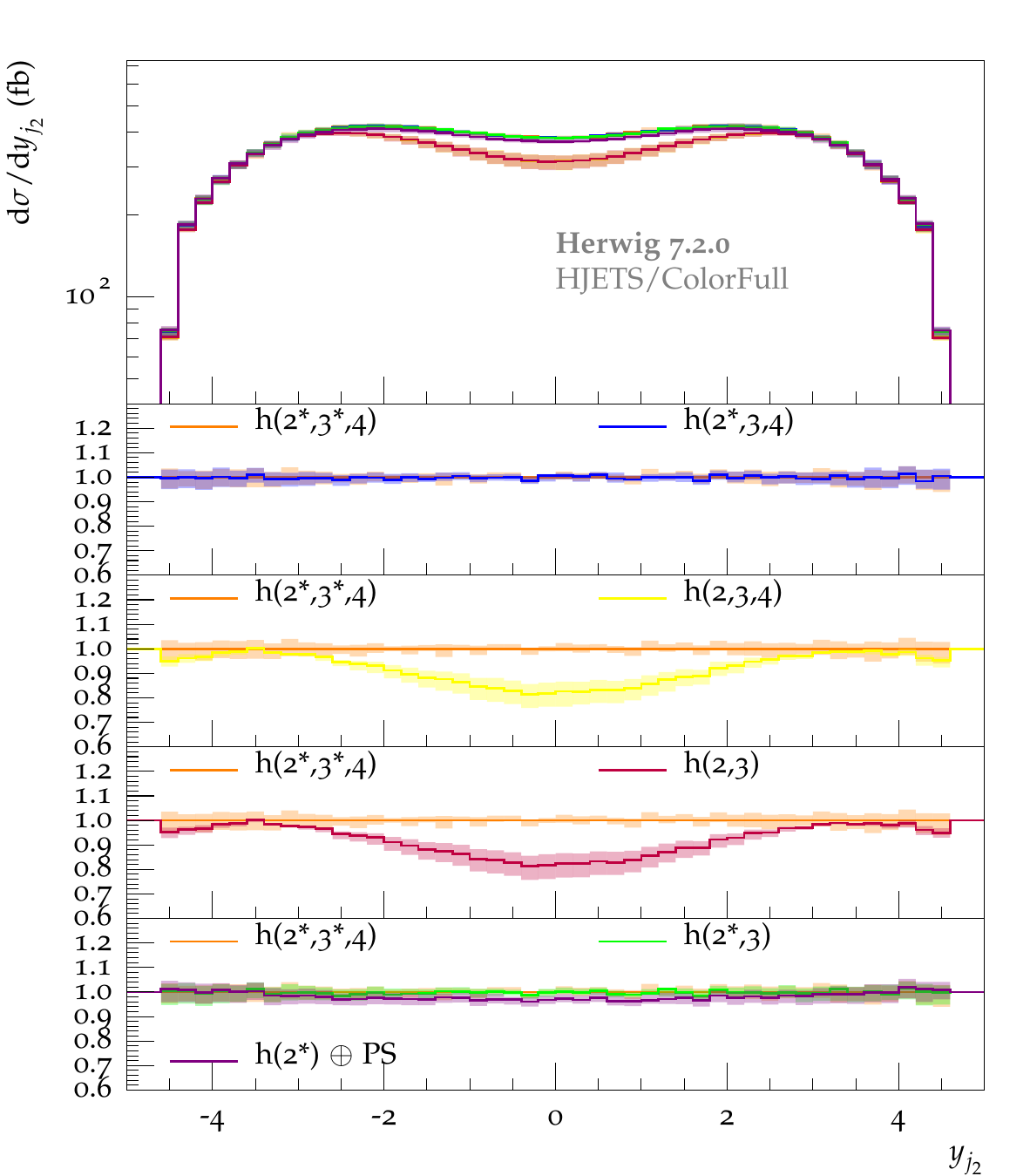} \\
        \includegraphics[width=0.45\textwidth]{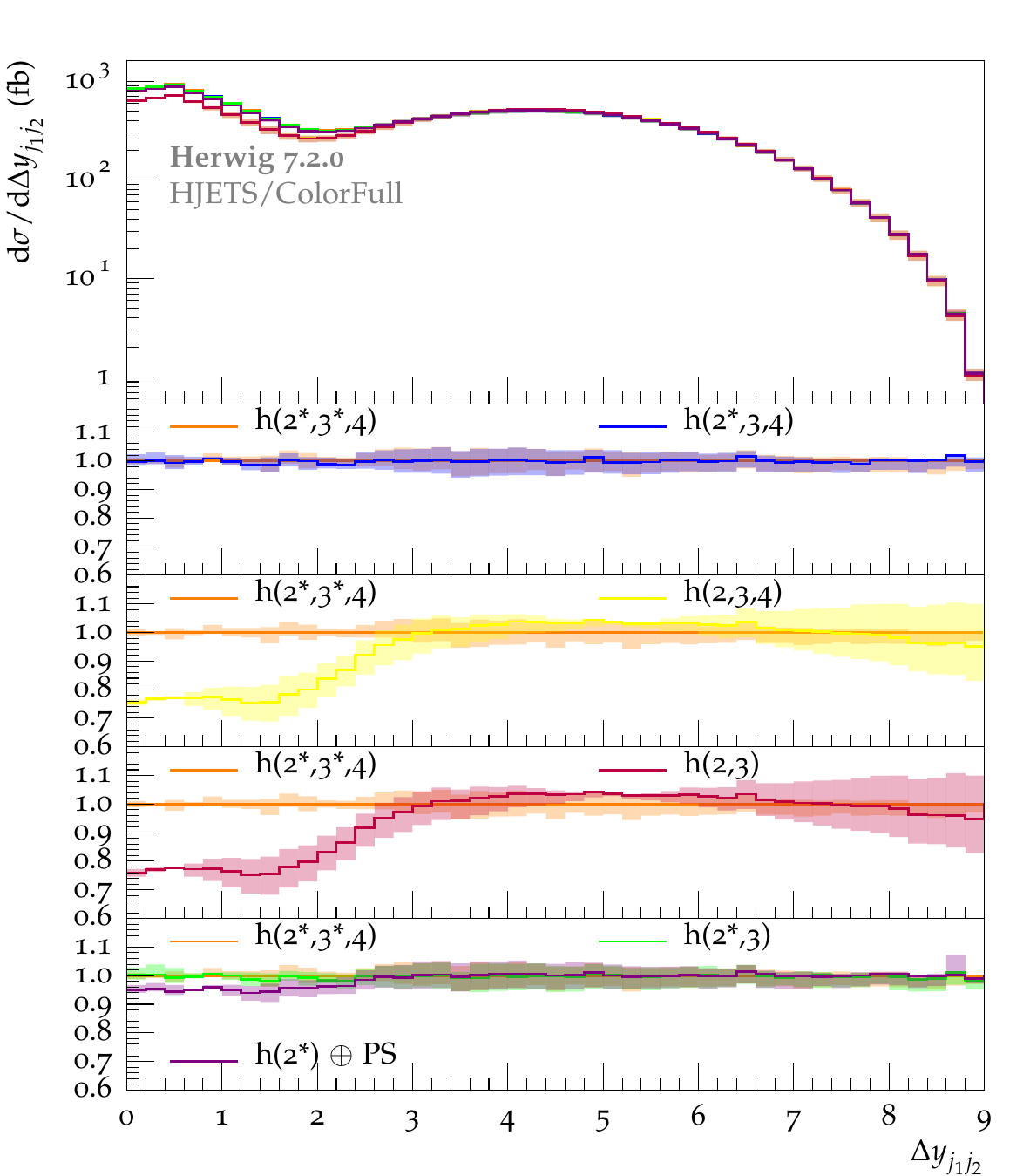} 
        \includegraphics[width=0.45\textwidth]{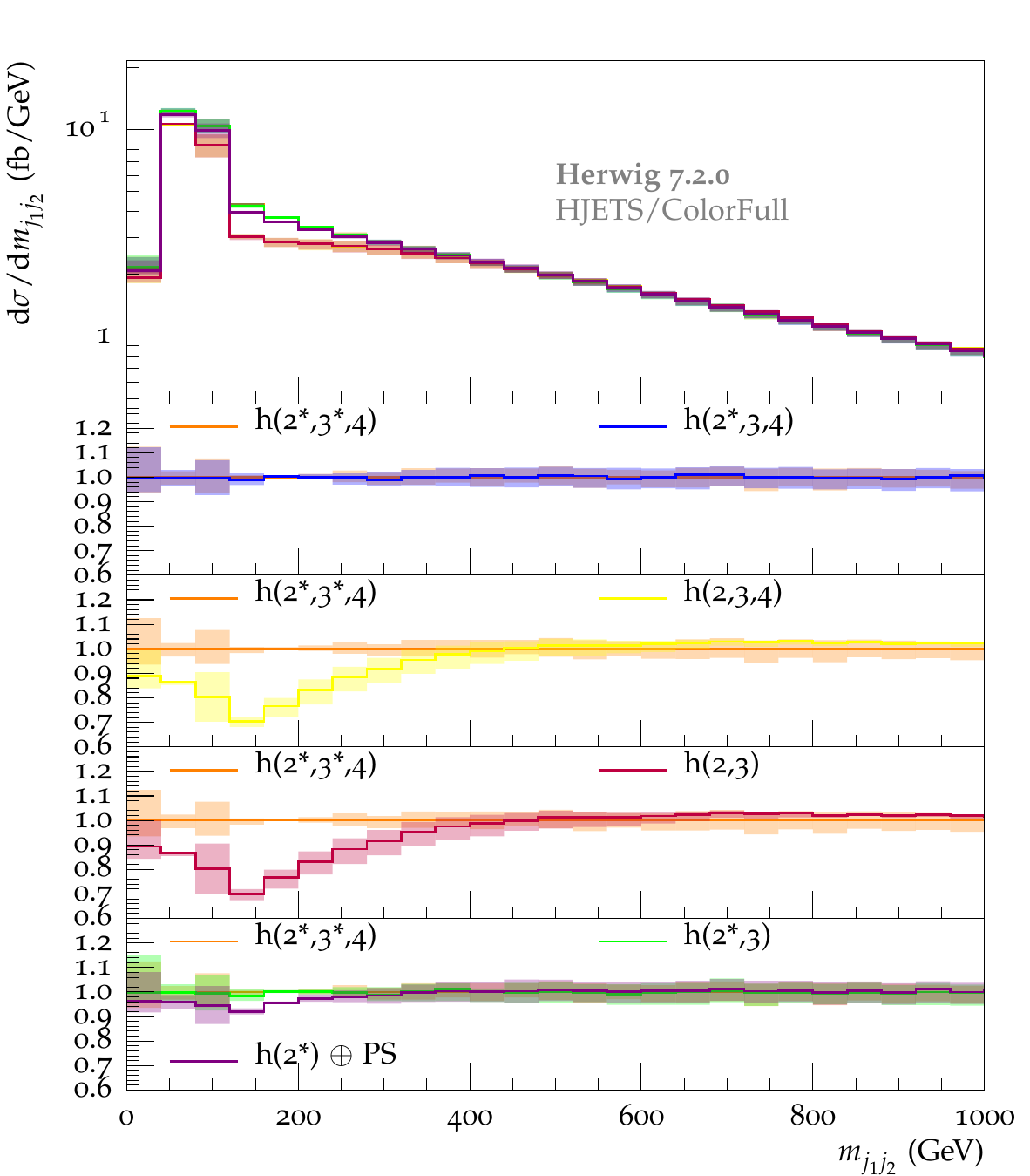}
    \caption{Shown is the distribution of 2 jets observables with INCL selection cuts: transverse momentum (top left) $p_{T,j_{2}}$ and rapidity $y_{j_{2}}$ (top right) for the second leading jet, rapidity gap $\Delta y_{j_{1}j_{2}}$(bottom left) and invariant mass $m_{j_{1}j_{2}}$ (bottom right) of the two tagging jets. The prediction is $h(2^{\star},3^{\star},4)$, $h(2^{\star},3,4)$, $h(2,3,4)$, $h(2^{\star},3)$, $h(2,3)$, and $h(2^{\star}) \bigoplus {\rm PS}$ with $h(2^{\star},3^{\star},4)$ being the reference in the ratio plots. Shown in the figures are theory error bands due to the variation of the factorization and renormalization scales.}
    \label{fig:2jets}
\end{figure*}

\section{Discussion of Results} %
\label{sec:results}
\subsection{Merging Scale Variations}
\label{sub:results-1}
In Fig.~\ref{fig:ktspl}, we show results for the splitting scales $\sqrt{d_{2}}$ and $\sqrt{d_{3}}$ of the $k_{T}$ jet algorithm \cite{Catani:1993hr,Ellis:1993tq}. The splitting scales are resolution scales $\sqrt{d_{k}}$ of the $k_{T}$ jet algorithm at which the event switches from $k$ jet event to a $k+1$ jet event. In the top row of Fig.~\ref{fig:ktspl} we show results for the merged setup $h(2^\star,3^\star,4)$ with the central merging scale $\rho_{C}$ set to the following values: $15,25,35$ GeV. In the transition region between the $h+n$ parton MEs and the $h+(n+1)$ parton MEs near $\sqrt{d_{n}} \approx 25$ GeV, we see $10\%$ variations for $n=2,3$. In the bottom row of Fig.~\ref{fig:ktspl} we compare the merged setups and matched setups against the merged setup $h(2^\star, 3^\star,4)$. The merging scale variations in all other observables we consider are much smaller than the other scale variations we consider and we hence do not show them in the plots to follow. The matched results \HjjnloPS and matching by merging setup $h(2^{\star},3)$ have a similar pattern after $\sqrt{d_{2}} \approx 125$ GeV and $\sqrt{d_{3}} \approx 125$ GeV. The matched results \HjjnloPS and \HjjjnloPS complement each other to some extent. 
The generation cut of $10$ GeV on transverse momentum for the \HjjjnloPS leads to a suppression around $\sqrt{d_{2}} \approx 10$ GeV and $\sqrt{d_{3}} \approx 10$ GeV. Furthermore, the matched \HjjjnloPS result is missing the $h+2$ jet events. The merged calculation is clearly reproduced by the matched ones in the regions where the respective jet multiplicity is resolved and the fixed order, hard jet multiplicity, provides a reliable prediction, however the merged description is able to interpolate in between the different multiplicities.
\subsection{Factorization and Renormalization Scale Variations}
\label{sub:fr}
In this subsection we present the results of simulations for the merged setups, $h(2^{\star},3^{\star},4)$, $h(2^{\star},3,4)$, $h(2^{\star},3)$, $h(2,3)$, $h(2,3,4)$ and the matched setups, \HjjnloPS~and \HjjjnloPS. Here we define the renormalization scale to be $\mu_{R}=\xi_{R} \mu_{0}$ and the factorization scale to be $\mu_{F}=\xi_{F} \mu_{0}$ with $\xi_{F}$ and $\xi_{R}$ denoting the scale factors. In the following figures the error bands are due to the variation of the renormalization and factorization scale factors $\xi_{F}$ and $\xi_{R}$ with $\xi_{F}=\xi_{R}=1/2,1$ and $2$. 


In the top row of Fig.~\ref{fig:xs} we show cross sections binned according to the exclusive number of identified jets $N_{\text{jets}}$ and to the exclusive number of identified gap jets $N_{\text{jets}}^{\text{gap}}$ for INCL selection cuts. 
In the ratio plots, all results are divided by the result of the merged setup $h(2^{\star},3^{\star},4)$. The matched setup \HjjnloPS when compared to the $h(2^{\star},3^{\star},4)$ agrees up until two identified jets and $0$ identified gap jets.  Further the \HjjnloPS result appears to underestimate the theoretical errors. In the bottom row of Fig.~\ref{fig:xs} we compare merged setups using the {\tt HJets} MEs and {\tt VBFNLO} MEs using TIGHT selection cuts. The reference in ratio plots is the prediction using the {\tt HJets} MEs. We see a good agreement for the binned cross section in the exclusive number of jets up to $4$ jets and for the binned cross section in the number of exclusive gap jets up to $2$ gap jets.  
The hard sub-processes using {\tt VBFNLO} and {\tt HJets} MEs are assigned different colour flows in the large-$N_{c}$ limit, since the full calculation takes into account all contributing topologies. Since the dipole shower evolution is intimately connected to the hard sub-process colour flow, different hard sub-process colour flows result in different shower histories.  The TIGHT selection cuts essentially allow extra jets in the gap between the leading two jets. Both $h+3$ and $h+4$ parton events allow for extra jets in the gap between the jets even after the strict TIGHT selection cuts. One should note that the rate is quite small in these higher multiplicity bins.

In Fig.~\ref{fig:2jets}, the distributions in transverse momentum and rapidity for the second leading jet are shown in addition to the kinematic distributions in the rapidity difference of the leading two jets $\Delta y_{j_{1}j_{2}}$ and the invariant mass of the leading two jets $m_{j_{1}j_{2}}$ are shown. We see good agreement between the matched setup \HjjnloPS and merged setups $h(2^{\star},3^{\star},4)$, $h(2^{\star},3,4)$ and $h(2^{\star},3)$ with the exception of a region between $100$ GeV and $150$ GeV in the invarant mass of leading two jets which is no more that $10 \%$. The leading order merged setups $h(2,3,4)$ and $h(2,3)$ deviate as expected.

\begin{figure*}
    \centering
        \includegraphics[width=0.45\textwidth]{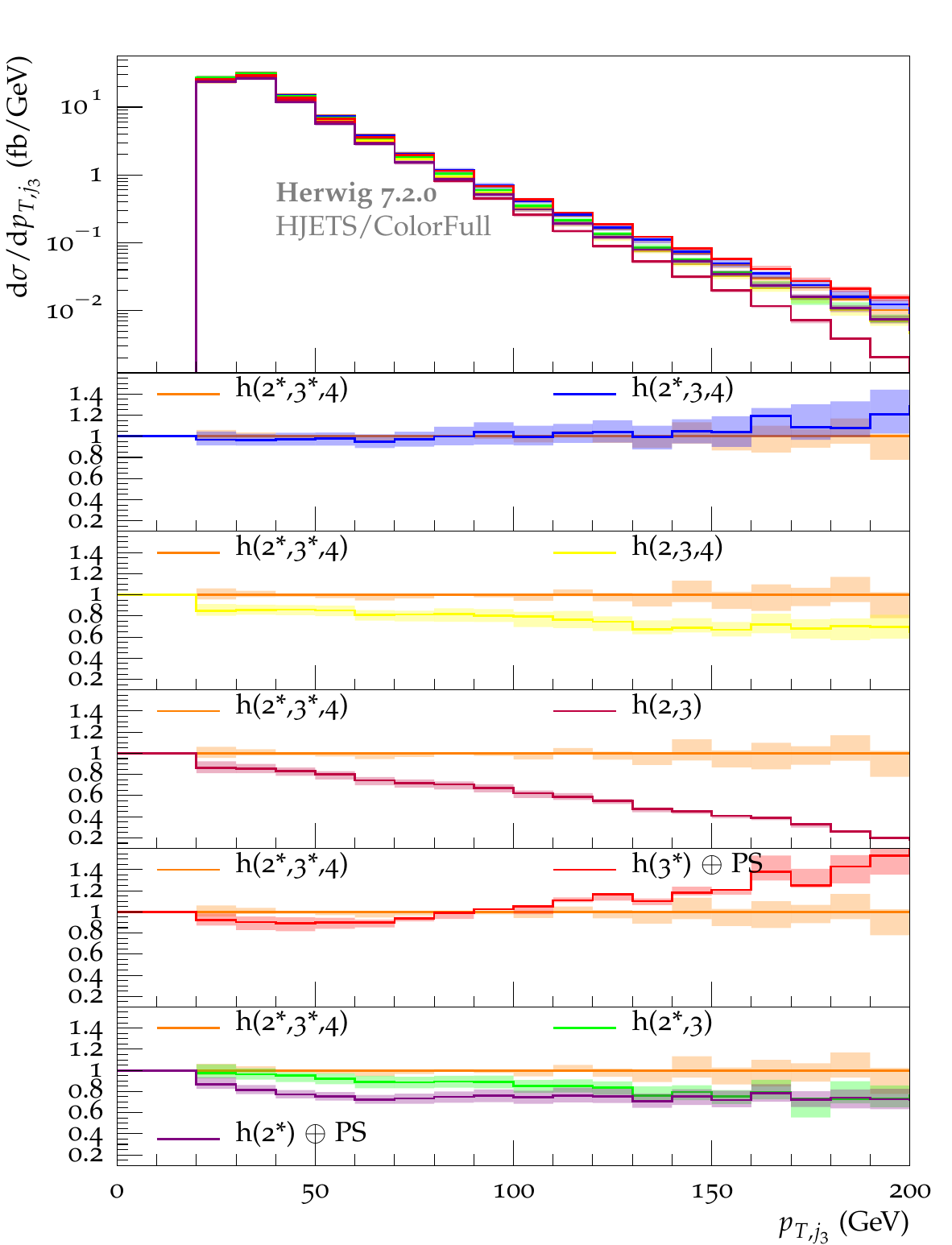} 
        \includegraphics[width=0.45\textwidth]{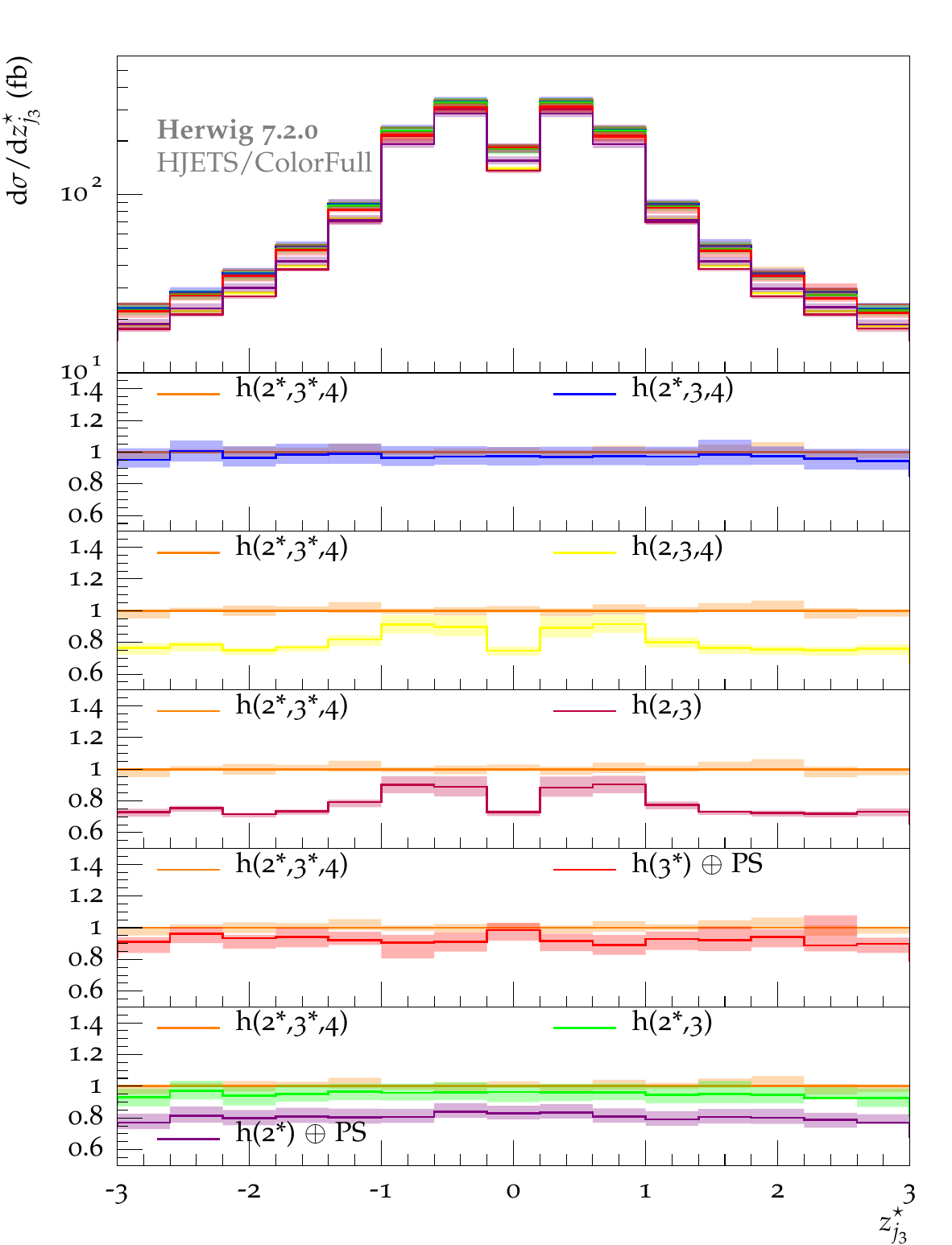} \\
        \includegraphics[width=0.45\textwidth]{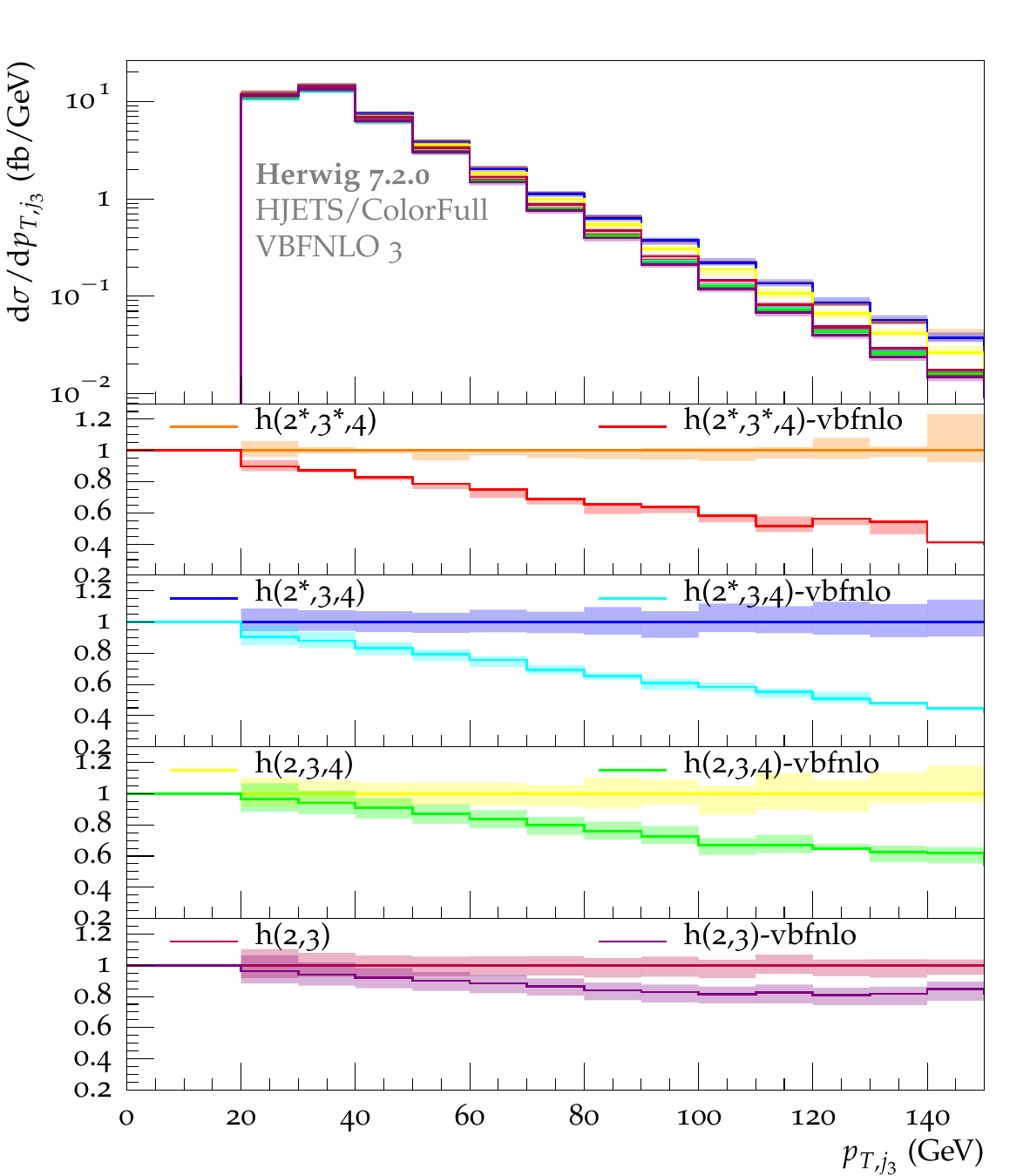}
        \includegraphics[width=0.45\textwidth]{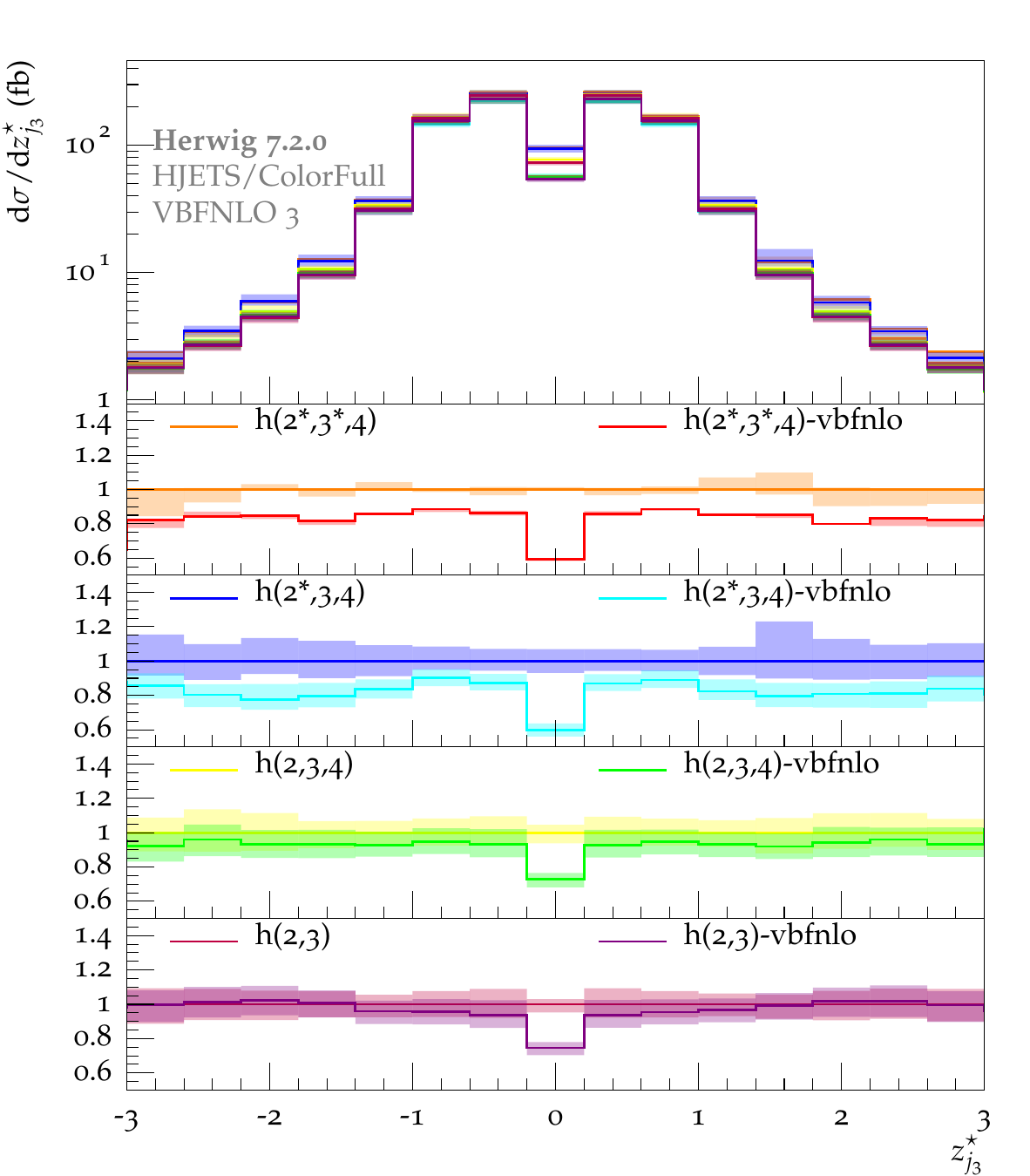}
    \caption{Shown are kinematic distributions in the transverse momentum of the third leading jet $p_{T,j_{3}}$ (left column) and Zeppenfeld variable $z^{\star}_{j_{3}}$ (right column). In the top row predictions $h(2^{\star},3^{\star},4)$, $h(2^{\star},3,4)$, $h(2,3,4)$, $h(2^{\star},3)$, $h(2,3)$, $h(3^{\star})\bigoplus {\rm PS}$, and $h(2^{\star}) \bigoplus {\rm PS}$ are compared to the $h(2^{\star},3^{\star},4)$ prediction (INCL selection cuts). In the bottom row we compare the merged setups using {\tt VBFNLO} MEs against the setups using {\tt HJets} MEs using LOOSE  selection cuts. Shown in the figures are theory error bands due to the variation of the factorization and renormalization scales. }
    \label{fig:jet3}
\end{figure*}
Shown in the top row of Fig.~\ref{fig:jet3} using INCL selection cuts are the kinematic distributions in the transverse momentum of third leading jet $p_{T,j_{3}}$ and the centrality of the third jet $z^{*}_{j_3}$ defined as 
\begin{equation}
    z^{*}_{j_3}=\frac{y_{j_3}-\frac{y_{j_1}+y_{j_2}}{2}}{|\Delta y_{j_1j_2}|}.
\end{equation}
For the transverse momentum and the centrality of the third jet, the matched setup \HjjnloPS when compared to $h(2^{\star},3^{\star},4)$ deviate by up to $20 \%$. This feature is shared by the matching by merging setups $h(2^{\star},3)$. It is quite obvious that relying on the parton shower alone without including the $h+4$ parton hard scattering matrix elements fails to describe the higher jet multiplicities beyond $2$. In the bottom row of Fig.~\ref{fig:jet3} using LOOSE selection cuts we see deviations between predictions based on {\tt VBFNLO} MEs and the {\tt HJets} MEs up to $50 \%$ in the tail of the distribution for the transverse momentum of the third jet for merged setup $h(2^\star,3^\star,4)$. For the centrality of the third jet, we see the largest deviations when using NLO MEs. All predictions deviate at the level $40\%$ around $z^{*}_{j_{3}}=0$. This is the result of the missing $s$-channel contributions in the {\tt VBFNLO} predictions. 
%

\begin{figure*}
    \centering
    \includegraphics[width=0.45\textwidth]{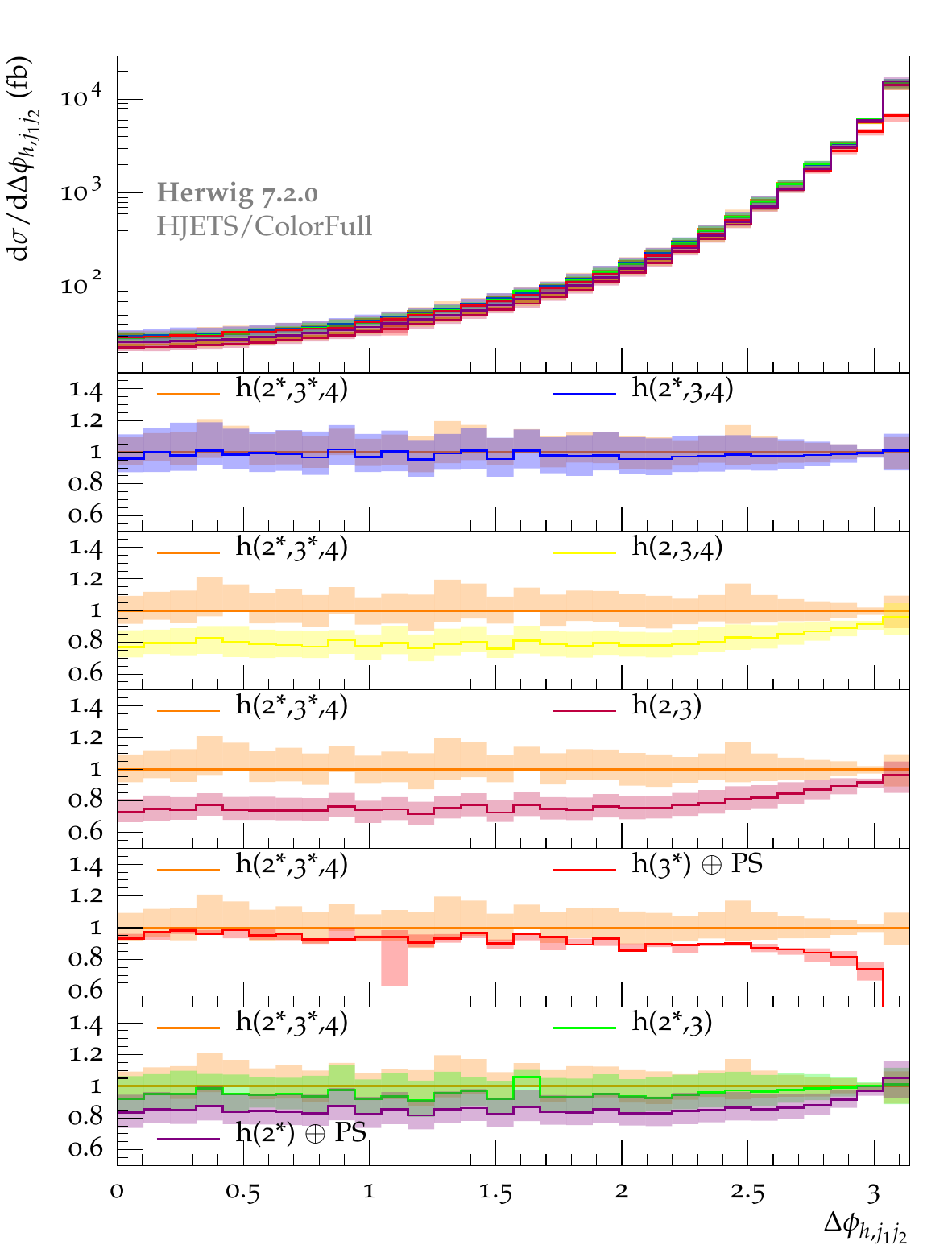} 
        \includegraphics[width=0.45\textwidth]{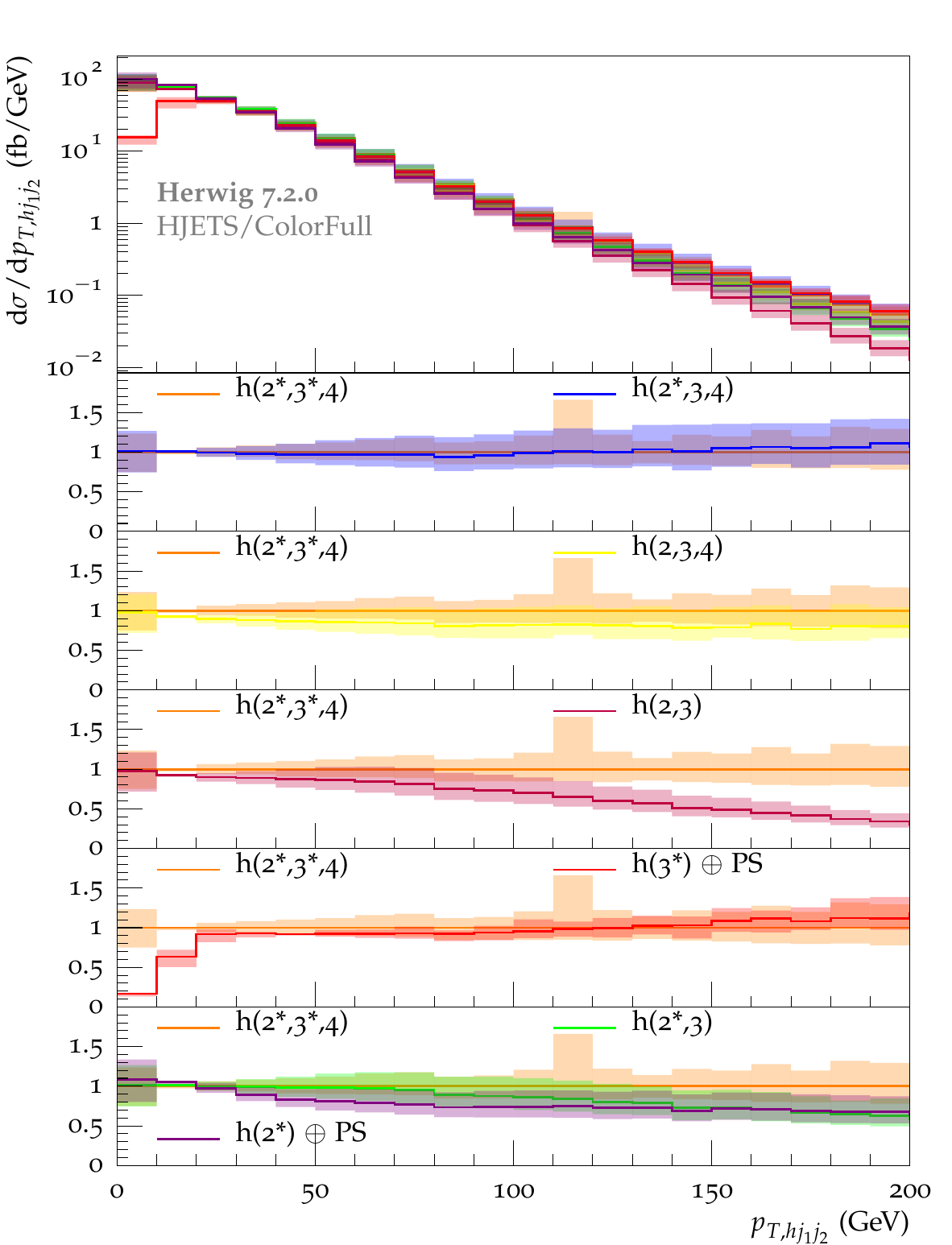} \\
        \includegraphics[width=0.45\textwidth]{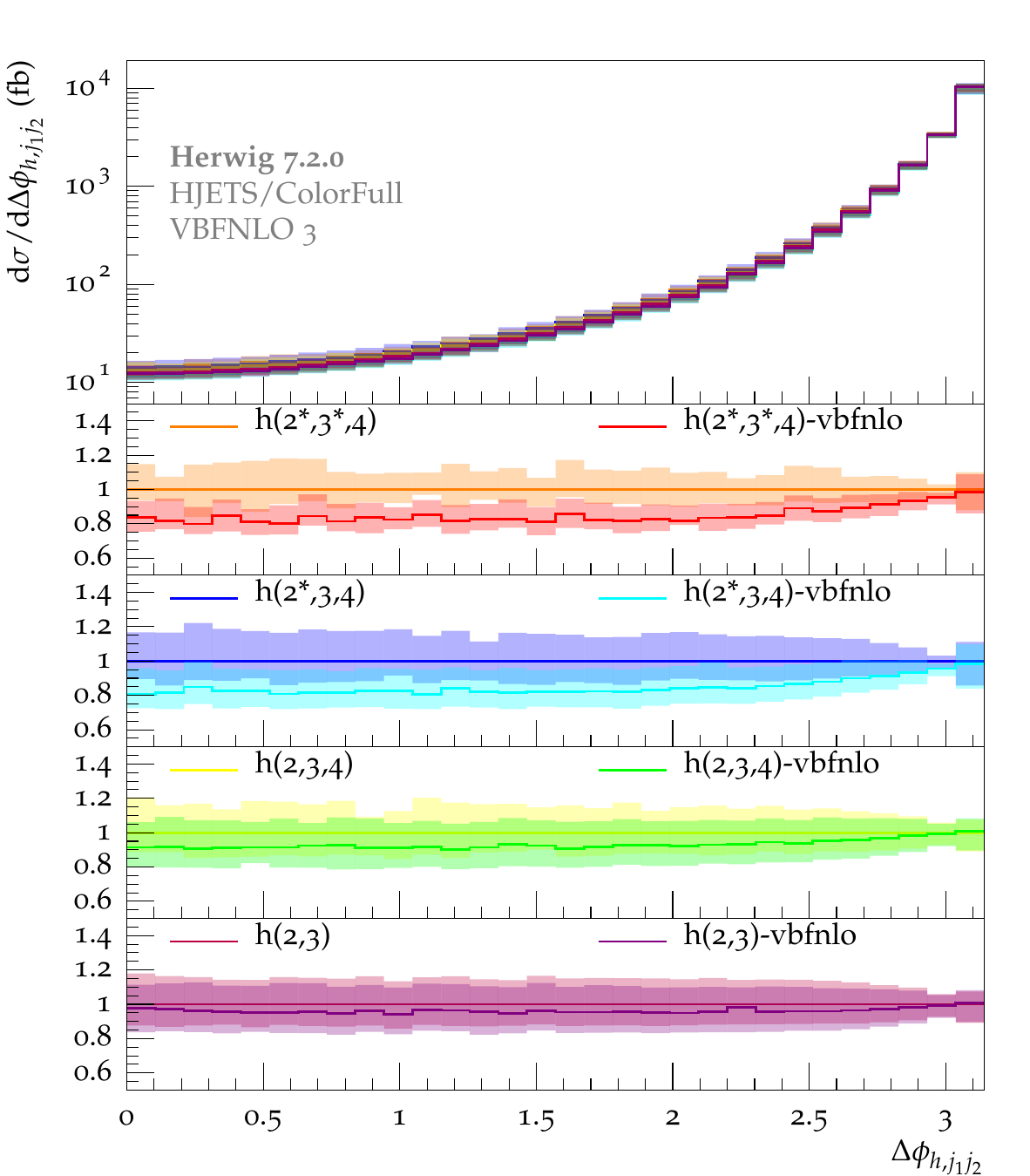}
        \includegraphics[width=0.45\textwidth]{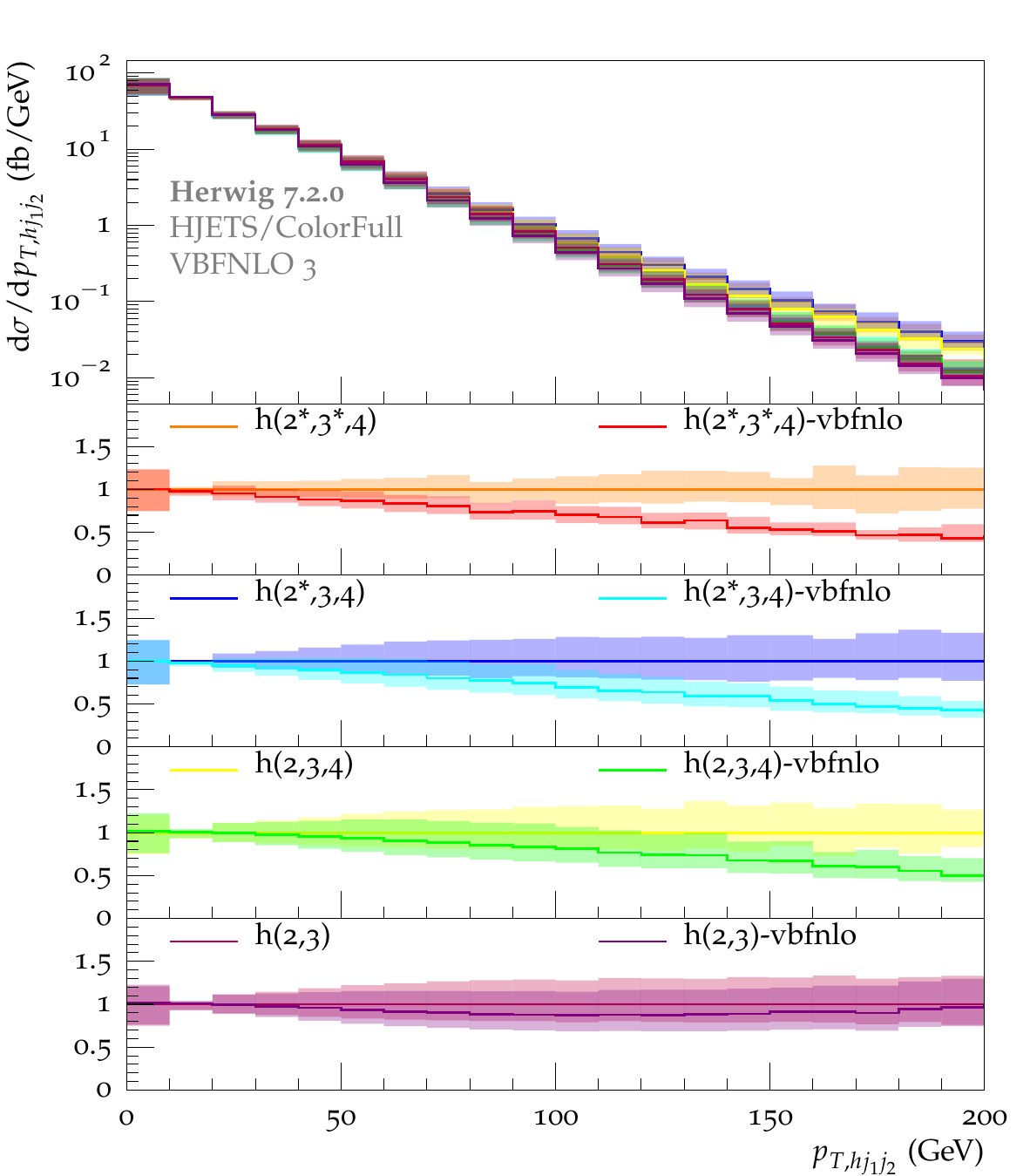}
    \caption{Shown are distributions in the azimuthal angle difference between the Higgs boson and system formed by the leading two jets $\Delta \phi_{h,j_{1}j_{2}}$ (left column) and in the transverse momentum of the Higgs boson plus two leading jets system $p_{T,hj_{1}j_{2}}$ (right column). In the top row predictions $h(2^{\star},3^{\star},4)$, $h(2^{\star},3,4)$, $h(2,3,4)$, $h(2^{\star},3)$, $h(2,3)$, $h(3^{\star})\bigoplus {\rm PS}$, and $h(2^{\star}) \bigoplus {\rm PS}$ are compared to the $h(2^{\star},3^{\star},4)$ prediction (INCL selection cuts). In the bottom row we compare the merged setups using {\tt VBFNLO} MEs against the setups using {\tt HJets} MEs for LOOSE selection cuts. Shown in the figures are theory error bands due to the variation of the factorization and renormalization scales.   }
    \label{fig:pthjj}
\end{figure*}
 
In Fig.~\ref{fig:pthjj} we show kinematic distributions for the azimuthal angle difference between the Higgs boson and two leading jets $\Delta \phi_{h,j_{1}j_{2}}$ and transverse momentum of the Higgs boson plus two leading jets $p_{T,hj_{1}j_{2}}$ that result from our matched and merged setups. The $\Delta \phi_{h,j_{1}j_{2}}$ observable is defined as 
\begin{equation}
    \Delta\phi_{h,j_{1}j_{2}}=|\phi_h-\phi_{j_1j_2}|,
\end{equation}
where $\phi_h$ is the azimuthal angle of the Higgs boson and $\phi_{j_1j_2}$ is the azimuthal angle of the system comprised of the leading two jets. The $p_{T,hj_{1}j_{2}}$ observable is defined by
\begin{equation}
    p_{T,hj_{1}j_{2}}=\big(p_h+p_{j_1}+p_{j_2}\big)_{T},
\end{equation}
where $p_h$, $p_{j_1}$, $p_{j_2}$ label the momentum of the Higgs boson, momentum of the first leading jet and the momentum of the second leading jet. The subscript $T$ represents the transverse component of the momentum sum.
For $\Delta \phi_{h,j_{1}j_{2}}$ we see an enhancement occurring in the region between $\Delta \phi_{h,j_{1}j_{2}}=0$ and $\Delta \phi_{h,j_{1}j_{2}}=2.9$ when comparing the matched \HjjnloPS results to that of the merged NLO $h(2^{\star},3^{\star},4)$ results. It should be noted that the NLO matched \HjjjnloPS result appears to underestimate the theory error associated with varying the factorization and renormalization scales. The distributions in $p_{T,hj_{1}j_{2}}$ shows an enhancement when comparing NLO matched result \HjjnloPS to the NLO merged $h(2^{\star},3^{\star},4)$ result. The theory error is underestimated by the NLO matched \HjjnloPS predictions. The matching by merging setups $h(2,3)$ and $h(2^{\star},3)$ follow a similar pattern to the \HjjnloPS setup. When comparing merged setups using the {\tt HJets} MEs against the {\tt VBFNLO} MEs for the LOOSE selection cuts we see deviations in the tail of the  $p_{T,hj_{1}j_{2}}$ distribution.  For $\Delta \phi_{h,j_{1}j_{2}}$ we notice when going from the LO MEs to NLO MEs the two predictions separate even further. This suggests that applying an overall $K$ factor to predictions using LO merged setup may not describe the physics very well.

\section{Conclusions and Outlook}
\label{sec:Conclusion}
For the first time NLO predictions for electroweak Higgs production have been merged with a dipole shower using the full set of tree-level and one-loop matrix elements available via the {\tt HJets} matrix element library. 
We have shown that for two jet observables that the merged setups $h(2^{\star},3^{\star},4)$ and matched setups \HjjnloPS are in good agreement. While for three jet observables such as the transverse momentum of the third hardest jet in the event, we note deviations as large as $20\%$ when comparing the merged setup $h(2^{\star},3^{\star},4)$ and matched setup \HjjnloPS. This can be attributed to the fact that the matched setup for this jet multiplicity is actually a leading order result. When comparing 
$h(2^{\star},3^{\star},4)$ and matched setup \HjjjnloPS we see deviations for most of the third jet observables. This can be viewed as follows: the matching uses a dynamic scale, but is fixed on the multijet topology, like the $H_{T,{\rm jets}}$ scale we have been using here. This also sets the hard shower scale. The merging in turn uses a more natural choice for multiscale problems, a fact which has extensively been discussed in the merging literature, see e.g. \cite{Bellm:2017ktr}, and is also at the heart of approaches like the MINLO algorithms \cite{Hamilton:2012np}. While for most of the jet distributions the matched and merged setups agree within 10-20\%, we conclude that the deviations for large $k_{T}$ splitting scales and third jet transverse momenta signal that fixed-multiplicity matching, even with a dynamic scale, might not be the appropriate simulation. We will investigate this issue in future work in more detail, in particular also their dependence on different jet radii.

Comparing the $h(2^{\star},3^{\star},4)$ merged setups using matrix elements from {\tt HJets} and {\tt VBFNLO} we see good agreement for the case of TIGHT selection cuts and strong deviations when using LOOSE selection cuts. As a best practice we recommend using {\tt HJets} when interested in a region of phase space described by the INCL and LOOSE selection cuts. 

With the level of control we have now gained through the multijet merging -- with and without the VBF approximation -- we are confident that algorithms come in reach which could utilize the properties of unitarized merging for the combination with NNLO. In this case one could systematically remove the approximated from the full calculations, and add them back using the available NNLO QCD corrections for the VH \cite{Ferrera:2013yga,Ferrera:2014lca,Ferrera:2014yda} and VBF-approximated channels~\cite{Cruz-Martinez:2018dvl,Cruz-Martinez:2018rod,Cacciari:2015jma}, respectively.

\begin{acknowledgement}
We thank Johannes Bellm for his early contributions to this project. All simulations presented in this paper were performed using the High Performance Computer (HPC) cluster {\tt Beocat} at Kansas State University and {\tt BeoShock} at Wichita State University. T.F and T.C would like to thank the Physics Division within the Department of Physics, Statistics, and Physics at Wichita State University (WSU) for supporting this project as well as WSU for supporting T.C. financially. This work has partly received funding from the European
Union’s Horizon 2020 research and innovation programme as part of the Marie Skłodowska-Curie Innovative Training Network MCnetITN3 (grant agreement no. 722104), and in part also by the COST actions CA16201 ``PARTICLEFACE'' and CA16108 ``VBSCAN''. SP
wants to thank the Erwin Schr\"odinger Institute Vienna for support
while this work has been finalized.
\end{acknowledgement}

\bibliographystyle{spphys}
\bibliography{vbf-merging}

\begin{thebibliography}{10}
\providecommand{\url}[1]{{#1}}
\providecommand{\urlprefix}{URL }
\expandafter\ifx\csname urlstyle\endcsname\relax
  \providecommand{\doi}[1]{DOI \discretionary{}{}{}#1}\else
  \providecommand{\doi}{DOI \discretionary{}{}{}\begingroup
  \urlstyle{rm}\Url}\fi

\bibitem{Chatrchyan:2012ufa}
S.~Chatrchyan, et~al., Phys. Lett. B \textbf{716}, 30 (2012).
\newblock \doi{10.1016/j.physletb.2012.08.021}

\bibitem{Aad:2012tfa}
G.~Aad, et~al., Phys. Lett. B \textbf{716}, 1 (2012).
\newblock \doi{10.1016/j.physletb.2012.08.020}

\bibitem{Aad:2013xqa}
G.~Aad, et~al., Phys. Lett. B \textbf{726}, 120 (2013).
\newblock \doi{10.1016/j.physletb.2013.08.026}

\bibitem{Chatrchyan:2012jja}
S.~Chatrchyan, et~al., Phys. Rev. Lett. \textbf{110}(8), 081803 (2013).
\newblock \doi{10.1103/PhysRevLett.110.081803}

\bibitem{Khachatryan:2014jba}
V.~Khachatryan, et~al., Eur. Phys. J. C \textbf{75}(5), 212 (2015).
\newblock \doi{10.1140/epjc/s10052-015-3351-7}

\bibitem{Kibble:1967sv}
T.~Kibble, Phys. Rev. \textbf{155}, 1554 (1967).
\newblock \doi{10.1103/PhysRev.155.1554}

\bibitem{Guralnik:1964eu}
G.~Guralnik, C.~Hagen, T.~Kibble, Phys. Rev. Lett. \textbf{13}, 585 (1964).
\newblock \doi{10.1103/PhysRevLett.13.585}

\bibitem{Englert:1964et}
F.~Englert, R.~Brout, Phys. Rev. Lett. \textbf{13}, 321 (1964).
\newblock \doi{10.1103/PhysRevLett.13.321}

\bibitem{Higgs:1964ia}
P.W. Higgs, Phys. Lett. \textbf{12}, 132 (1964).
\newblock \doi{10.1016/0031-9163(64)91136-9}

\bibitem{Higgs:1964pj}
P.W. Higgs, Phys. Rev. Lett. \textbf{13}, 508 (1964).
\newblock \doi{10.1103/PhysRevLett.13.508}

\bibitem{Higgs:1966ev}
P.W. Higgs, Phys. Rev. \textbf{145}, 1156 (1966).
\newblock \doi{10.1103/PhysRev.145.1156}

\bibitem{Weinberg:1967tq}
S.~Weinberg, Phys. Rev. Lett. \textbf{19}, 1264 (1967).
\newblock \doi{10.1103/PhysRevLett.19.1264}

\bibitem{Salam:1968rm}
A.~Salam, Conf. Proc. C \textbf{680519}, 367 (1968).
\newblock \doi{10.1142/9789812795915_0034}

\bibitem{Glashow:1961tr}
S.~Glashow, Nucl. Phys. \textbf{22}, 579 (1961).
\newblock \doi{10.1016/0029-5582(61)90469-2}

\bibitem{Sirunyan:2021ybb}
A.M. Sirunyan, et~al.,   (2021)

\bibitem{Sirunyan:2021rug}
A.M. Sirunyan, et~al.,   (2021)

\bibitem{Sirunyan:2020tzo}
A.M. Sirunyan, et~al., JHEP \textbf{03}, 003 (2021).
\newblock \doi{10.1007/JHEP03(2021)003}

\bibitem{Sirunyan:2018egh}
A.M. Sirunyan, et~al., Phys. Lett. B \textbf{791}, 96 (2019).
\newblock \doi{10.1016/j.physletb.2018.12.073}

\bibitem{Sirunyan:2017tqd}
A.M. Sirunyan, et~al., Phys. Lett. B \textbf{775}, 1 (2017).
\newblock \doi{10.1016/j.physletb.2017.10.021}

\bibitem{Sirunyan:2017khh}
A.M. Sirunyan, et~al., Phys. Lett. B \textbf{779}, 283 (2018).
\newblock \doi{10.1016/j.physletb.2018.02.004}

\bibitem{Sirunyan:2018ouh}
A.M. Sirunyan, et~al., JHEP \textbf{11}, 185 (2018).
\newblock \doi{10.1007/JHEP11(2018)185}

\bibitem{Sirunyan:2018koj}
A.M. Sirunyan, et~al., Eur. Phys. J. C \textbf{79}(5), 421 (2019).
\newblock \doi{10.1140/epjc/s10052-019-6909-y}

\bibitem{Sirunyan:2019nbs}
A.M. Sirunyan, et~al., Phys. Rev. D \textbf{100}(11), 112002 (2019).
\newblock \doi{10.1103/PhysRevD.100.112002}

\bibitem{Aad:2020mkp}
G.~Aad, et~al., Eur. Phys. J. C \textbf{80}(10), 957 (2020).
\newblock \doi{10.1140/epjc/s10052-020-8227-9}

\bibitem{Aad:2020mnm}
G.~Aad, et~al., Phys. Lett. B \textbf{805}, 135426 (2020).
\newblock \doi{10.1016/j.physletb.2020.135426}

\bibitem{CMS:2019jdw}
A.M. Sirunyan, et~al., Phys. Rev. D \textbf{100}(11), 112002 (2019).
\newblock \doi{10.1103/PhysRevD.100.112002}

\bibitem{ATLAS-CONF-2021-014}
{Measurements of gluon fusion and vector-boson-fusion production of the Higgs
  boson in $H\rightarrow W W^* \rightarrow e\nu \mu\nu$ decays using $pp$
  collisions at $\sqrt{s}=13$ TeV with the ATLAS detector}.
\newblock Tech. rep., CERN, Geneva (2021).
\newblock \urlprefix\url{https://cds.cern.ch/record/2759651}.
\newblock All figures including auxiliary figures are available at
  https://atlas.web.cern.ch/Atlas/GROUPS/PHYSICS/CONFNOTES/ATLAS-CONF-2021-014

\bibitem{ATLAS:2020wny}
G.~Aad, et~al., Eur. Phys. J. C \textbf{80}(10), 942 (2020).
\newblock \doi{10.1140/epjc/s10052-020-8223-0}

\bibitem{CMS:2020dvg}
A.M. Sirunyan, et~al., JHEP \textbf{03}, 003 (2021).
\newblock \doi{10.1007/JHEP03(2021)003}

\bibitem{Han:1992hr}
T.~Han, G.~Valencia, S.~Willenbrock, Phys. Rev. Lett. \textbf{69}, 3274 (1992).
\newblock \doi{10.1103/PhysRevLett.69.3274}

\bibitem{Bolzoni:2010xr}
P.~Bolzoni, F.~Maltoni, S.O. Moch, M.~Zaro, Phys. Rev. Lett. \textbf{105},
  011801 (2010).
\newblock \doi{10.1103/PhysRevLett.105.011801}

\bibitem{Dreyer:2016oyx}
F.A. Dreyer, A.~Karlberg, Phys. Rev. Lett. \textbf{117}(7), 072001 (2016).
\newblock \doi{10.1103/PhysRevLett.117.072001}

\bibitem{Figy:2003nv}
T.~Figy, C.~Oleari, D.~Zeppenfeld, Phys. Rev. D \textbf{68}, 073005 (2003).
\newblock \doi{10.1103/PhysRevD.68.073005}

\bibitem{Berger:2004pca}
E.L. Berger, J.M. Campbell, Phys. Rev. D \textbf{70}, 073011 (2004).
\newblock \doi{10.1103/PhysRevD.70.073011}

\bibitem{Ciccolini:2007ec}
M.~Ciccolini, A.~Denner, S.~Dittmaier, Phys. Rev. D \textbf{77}, 013002 (2008).
\newblock \doi{10.1103/PhysRevD.77.013002}

\bibitem{Cacciari:2015jma}
M.~Cacciari, F.A. Dreyer, A.~Karlberg, G.P. Salam, G.~Zanderighi, Phys. Rev.
  Lett. \textbf{115}(8), 082002 (2015).
\newblock \doi{10.1103/PhysRevLett.115.082002}.
\newblock [Erratum: Phys.Rev.Lett. 120, 139901 (2018)]

\bibitem{Cruz-Martinez:2018dvl}
J.~Cruz-Martinez, E.W.N. Glover, T.~Gehrmann, A.~Huss, PoS \textbf{LL2018}, 003
  (2018).
\newblock \doi{10.22323/1.303.0003}

\bibitem{Cruz-Martinez:2018rod}
J.~Cruz-Martinez, T.~Gehrmann, E.W.N. Glover, A.~Huss, Phys. Lett. B
  \textbf{781}, 672 (2018).
\newblock \doi{10.1016/j.physletb.2018.04.046}

\bibitem{Figy:2007kv}
T.~Figy, V.~Hankele, D.~Zeppenfeld, JHEP \textbf{02}, 076 (2008).
\newblock \doi{10.1088/1126-6708/2008/02/076}

\bibitem{Jager:2020hkz}
B.~J\"ager, A.~Karlberg, S.~Pl\"atzer, J.~Scheller, M.~Zaro, Eur. Phys. J. C
  \textbf{80}(8), 756 (2020).
\newblock \doi{10.1140/epjc/s10052-020-8326-7}

\bibitem{Nason:2009ai}
P.~Nason, C.~Oleari, JHEP \textbf{02}, 037 (2010).
\newblock \doi{10.1007/JHEP02(2010)037}

\bibitem{Frixione:2013mta}
S.~Frixione, P.~Torrielli, M.~Zaro, Phys. Lett. B \textbf{726}, 273 (2013).
\newblock \doi{10.1016/j.physletb.2013.08.030}

\bibitem{DErrico:2011wfa}
L.~D'Errico, P.~Richardson, Eur. Phys. J. C \textbf{72}, 2042 (2012).
\newblock \doi{10.1140/epjc/s10052-012-2042-x}

\bibitem{Azzi:2019yne}
P.~Azzi, et~al., CERN Yellow Rep. Monogr. \textbf{7}, 1 (2019).
\newblock \doi{10.23731/CYRM-2019-007.1}

\bibitem{Jager:2014vna}
B.~J\"ager, F.~Schissler, D.~Zeppenfeld, JHEP \textbf{07}, 125 (2014).
\newblock \doi{10.1007/JHEP07(2014)125}

\bibitem{deFlorian:2016spz}
D.~de~Florian, et~al.,  \textbf{2/2017} (2016).
\newblock \doi{10.23731/CYRM-2017-002}

\bibitem{Hoche:2021mkv}
S.~H\"oche, S.~Mrenna, S.~Payne, C.T. Preuss, P.~Skands,   (2021)

\bibitem{Campanario:2013fsa}
F.~Campanario, T.M. Figy, S.~Plätzer, M.~Sjödahl, Phys. Rev. Lett.
  \textbf{111}(21), 211802 (2013).
\newblock \doi{10.1103/PhysRevLett.111.211802}

\bibitem{Campanario:2018ppz}
F.~Campanario, T.M. Figy, S.~Plätzer, M.~Rauch, P.~Schichtel, M.~Sjödahl,
  Phys. Rev. \textbf{D98}(3), 033003 (2018).
\newblock \doi{10.1103/PhysRevD.98.033003}

\bibitem{Liu:2019tuy}
T.~Liu, K.~Melnikov, A.A. Penin, Phys. Rev. Lett. \textbf{123}(12), 122002
  (2019).
\newblock \doi{10.1103/PhysRevLett.123.122002}

\bibitem{Dreyer:2020urf}
F.A. Dreyer, A.~Karlberg, L.~Tancredi, JHEP \textbf{10}, 131 (2020).
\newblock \doi{10.1007/JHEP10(2020)131}

\bibitem{Bahr:2008pv}
M.~Bahr, et~al., Eur. Phys. J. C \textbf{58}, 639 (2008).
\newblock \doi{10.1140/epjc/s10052-008-0798-9}

\bibitem{Bellm:2015jjp}
J.~Bellm, et~al., Eur. Phys. J. C \textbf{76}(4), 196 (2016).
\newblock \doi{10.1140/epjc/s10052-016-4018-8}

\bibitem{Platzer:2011bc}
S.~Platzer, S.~Gieseke, Eur. Phys. J. C \textbf{72}, 2187 (2012).
\newblock \doi{10.1140/epjc/s10052-012-2187-7}

\bibitem{Bellm:2019zci}
J.~Bellm, et~al., Eur. Phys. J. C \textbf{80}(5), 452 (2020).
\newblock \doi{10.1140/epjc/s10052-020-8011-x}

\bibitem{Campanario:2013nca}
F.~Campanario, T.M. Figy, S.~Plätzer, M.~Sjödahl, PoS \textbf{RADCOR2013},
  042 (2013).
\newblock \doi{10.22323/1.197.0042}

\bibitem{Campanario:2014aia}
F.~Campanario, T.M. Figy, S.~Plätzer, M.~Sjodahl, PoS \textbf{LL2014}, 025
  (2014).
\newblock \doi{10.22323/1.211.0025}

\bibitem{Sjodahl:2014opa}
M.~Sjodahl, Eur. Phys. J. \textbf{C75}(5), 236 (2015).
\newblock \doi{10.1140/epjc/s10052-015-3417-6}

\bibitem{Campanario:2011cs}
F.~Campanario, JHEP \textbf{10}, 070 (2011).
\newblock \doi{10.1007/JHEP10(2011)070}

\bibitem{Baglio:2014uba}
J.~Baglio, et~al.,   (2014)

\bibitem{Arnold:2011wj}
J.~Baglio, et~al.,   (2011)

\bibitem{Arnold:2008rz}
K.~Arnold, et~al., Comput. Phys. Commun. \textbf{180}, 1661 (2009).
\newblock \doi{10.1016/j.cpc.2009.03.006}

\bibitem{Alioli:2013nda}
S.~Alioli, et~al., Comput. Phys. Commun. \textbf{185}, 560 (2014).
\newblock \doi{10.1016/j.cpc.2013.10.020}

\bibitem{Binoth:2010xt}
T.~Binoth, et~al., Comput. Phys. Commun. \textbf{181}, 1612 (2010).
\newblock \doi{10.1016/j.cpc.2010.05.016}

\bibitem{Butterworth_2016}
J.~Butterworth, S.~Carrazza, A.~Cooper-Sarkar, A.D. Roeck, J.~Feltesse,
  S.~Forte, J.~Gao, S.~Glazov, J.~Huston, Z.~Kassabov, et~al., Journal of
  Physics G: Nuclear and Particle Physics \textbf{43}(2), 023001 (2016).
\newblock \doi{10.1088/0954-3899/43/2/023001}.
\newblock \urlprefix\url{http://dx.doi.org/10.1088/0954-3899/43/2/023001}

\bibitem{Cacciari:2005hq}
M.~Cacciari, G.P. Salam, Phys. Lett. \textbf{B641}, 57 (2006).
\newblock \doi{10.1016/j.physletb.2006.08.037}

\bibitem{Cacciari:2011ma}
M.~Cacciari, G.P. Salam, G.~Soyez, Eur. Phys. J. \textbf{C72}, 1896 (2012).
\newblock \doi{10.1140/epjc/s10052-012-1896-2}

\bibitem{Buckley_2013}
A.~Buckley, J.~Butterworth, D.~Grellscheid, H.~Hoeth, L.~Lönnblad, J.~Monk,
  H.~Schulz, F.~Siegert, Computer Physics Communications \textbf{184}(12),
  2803–2819 (2013).
\newblock \doi{10.1016/j.cpc.2013.05.021}.
\newblock \urlprefix\url{http://dx.doi.org/10.1016/j.cpc.2013.05.021}

\bibitem{Platzer:2009jq}
S.~Platzer, S.~Gieseke, JHEP \textbf{01}, 024 (2011).
\newblock \doi{10.1007/JHEP01(2011)024}

\bibitem{Frixione:2002ik}
S.~Frixione, B.R. Webber, JHEP \textbf{06}, 029 (2002).
\newblock \doi{10.1088/1126-6708/2002/06/029}

\bibitem{Nason:2004rx}
P.~Nason, JHEP \textbf{11}, 040 (2004).
\newblock \doi{10.1088/1126-6708/2004/11/040}

\bibitem{Alwall:2007fs}
J.~Alwall, et~al., Eur. Phys. J. C \textbf{53}, 473 (2008).
\newblock \doi{10.1140/epjc/s10052-007-0490-5}

\bibitem{Bellm:2017ktr}
J.~Bellm, S.~Gieseke, S.~Pl\"atzer, Eur. Phys. J. C \textbf{78}(3), 244 (2018).
\newblock \doi{10.1140/epjc/s10052-018-5723-2}

\bibitem{Catani:1990rr}
S.~Catani, B.~Webber, G.~Marchesini, Nucl. Phys. B \textbf{349}, 635 (1991).
\newblock \doi{10.1016/0550-3213(91)90390-J}

\end{thebibliography}


\begin{thebibliography}{10}
\providecommand{\url}[1]{{#1}}
\providecommand{\urlprefix}{URL }
\expandafter\ifx\csname urlstyle\endcsname\relax
  \providecommand{\doi}[1]{DOI \discretionary{}{}{}#1}\else
  \providecommand{\doi}{DOI \discretionary{}{}{}\begingroup
  \urlstyle{rm}\Url}\fi

\bibitem{Weinberg:1967tq}
S.~Weinberg, Phys. Rev. Lett. \textbf{19}, 1264 (1967).
\newblock \doi{10.1103/PhysRevLett.19.1264}

\bibitem{Salam:1968rm}
A.~Salam, Conf. Proc. C \textbf{680519}, 367 (1968).
\newblock \doi{10.1142/9789812795915_0034}

\bibitem{Glashow:1961tr}
S.~Glashow, Nucl. Phys. \textbf{22}, 579 (1961).
\newblock \doi{10.1016/0029-5582(61)90469-2}

\bibitem{Kibble:1967sv}
T.~Kibble, Phys. Rev. \textbf{155}, 1554 (1967).
\newblock \doi{10.1103/PhysRev.155.1554}

\bibitem{Guralnik:1964eu}
G.~Guralnik, C.~Hagen, T.~Kibble, Phys. Rev. Lett. \textbf{13}, 585 (1964).
\newblock \doi{10.1103/PhysRevLett.13.585}

\bibitem{Englert:1964et}
F.~Englert, R.~Brout, Phys. Rev. Lett. \textbf{13}, 321 (1964).
\newblock \doi{10.1103/PhysRevLett.13.321}

\bibitem{Higgs:1964ia}
P.W. Higgs, Phys. Lett. \textbf{12}, 132 (1964).
\newblock \doi{10.1016/0031-9163(64)91136-9}

\bibitem{Higgs:1964pj}
P.W. Higgs, Phys. Rev. Lett. \textbf{13}, 508 (1964).
\newblock \doi{10.1103/PhysRevLett.13.508}

\bibitem{Higgs:1966ev}
P.W. Higgs, Phys. Rev. \textbf{145}, 1156 (1966).
\newblock \doi{10.1103/PhysRev.145.1156}

\bibitem{Kauer:2000hi}
N.~Kauer, T.~Plehn, D.L. Rainwater, D.~Zeppenfeld, Phys. Lett. B \textbf{503},
  113 (2001).
\newblock \doi{10.1016/S0370-2693(01)00211-8}

\bibitem{Rainwater:1997dg}
D.L. Rainwater, D.~Zeppenfeld, JHEP \textbf{12}, 005 (1997).
\newblock \doi{10.1088/1126-6708/1997/12/005}

\bibitem{Rainwater:1998kj}
D.L. Rainwater, D.~Zeppenfeld, K.~Hagiwara, Phys. Rev. D \textbf{59}, 014037
  (1998).
\newblock \doi{10.1103/PhysRevD.59.014037}

\bibitem{Rainwater:1999sd}
D.L. Rainwater, D.~Zeppenfeld, Phys. Rev. D \textbf{60}, 113004 (1999).
\newblock \doi{10.1103/PhysRevD.60.113004}.
\newblock [Erratum: Phys.Rev.D 61, 099901 (2000)]

\bibitem{Asai:2004ws}
S.~Asai, et~al., Eur. Phys. J. C \textbf{32S2}, 19 (2004).
\newblock \doi{10.1140/epjcd/s2003-01-010-8}

\bibitem{Cranmer:2004uz}
K.~Cranmer, B.~Mellado, W.~Quayle, S.L. Wu,   (2004)

\bibitem{Eboli:2000ze}
O.J.P. Eboli, D.~Zeppenfeld, Phys. Lett. B \textbf{495}, 147 (2000).
\newblock \doi{10.1016/S0370-2693(00)01213-2}

\bibitem{Ciccolini:2007ec}
M.~Ciccolini, A.~Denner, S.~Dittmaier, Phys. Rev. D \textbf{77}, 013002 (2008).
\newblock \doi{10.1103/PhysRevD.77.013002}

\bibitem{Han:1992hr}
T.~Han, G.~Valencia, S.~Willenbrock, Phys. Rev. Lett. \textbf{69}, 3274 (1992).
\newblock \doi{10.1103/PhysRevLett.69.3274}

\bibitem{Bolzoni:2010xr}
P.~Bolzoni, F.~Maltoni, S.O. Moch, M.~Zaro, Phys. Rev. Lett. \textbf{105},
  011801 (2010).
\newblock \doi{10.1103/PhysRevLett.105.011801}

\bibitem{Dreyer:2016oyx}
F.A. Dreyer, A.~Karlberg, Phys. Rev. Lett. \textbf{117}(7), 072001 (2016).
\newblock \doi{10.1103/PhysRevLett.117.072001}

\bibitem{Figy:2003nv}
T.~Figy, C.~Oleari, D.~Zeppenfeld, Phys. Rev. D \textbf{68}, 073005 (2003).
\newblock \doi{10.1103/PhysRevD.68.073005}

\bibitem{Berger:2004pca}
E.L. Berger, J.M. Campbell, Phys. Rev. D \textbf{70}, 073011 (2004).
\newblock \doi{10.1103/PhysRevD.70.073011}

\bibitem{Cacciari:2015jma}
M.~Cacciari, F.A. Dreyer, A.~Karlberg, G.P. Salam, G.~Zanderighi, Phys. Rev.
  Lett. \textbf{115}(8), 082002 (2015).
\newblock \doi{10.1103/PhysRevLett.115.082002}.
\newblock [Erratum: Phys.Rev.Lett. 120, 139901 (2018)]

\bibitem{Cruz-Martinez:2018dvl}
J.~Cruz-Martinez, E.W.N. Glover, T.~Gehrmann, A.~Huss, PoS \textbf{LL2018}, 003
  (2018).
\newblock \doi{10.22323/1.303.0003}

\bibitem{Cruz-Martinez:2018rod}
J.~Cruz-Martinez, T.~Gehrmann, E.W.N. Glover, A.~Huss, Phys. Lett. B
  \textbf{781}, 672 (2018).
\newblock \doi{10.1016/j.physletb.2018.04.046}

\bibitem{Figy:2007kv}
T.~Figy, V.~Hankele, D.~Zeppenfeld, JHEP \textbf{02}, 076 (2008).
\newblock \doi{10.1088/1126-6708/2008/02/076}

\bibitem{Jager:2020hkz}
B.~J\"ager, A.~Karlberg, S.~Pl\"atzer, J.~Scheller, M.~Zaro, Eur. Phys. J. C
  \textbf{80}(8), 756 (2020).
\newblock \doi{10.1140/epjc/s10052-020-8326-7}

\bibitem{Nason:2009ai}
P.~Nason, C.~Oleari, JHEP \textbf{02}, 037 (2010).
\newblock \doi{10.1007/JHEP02(2010)037}

\bibitem{Frixione:2013mta}
S.~Frixione, P.~Torrielli, M.~Zaro, Phys. Lett. B \textbf{726}, 273 (2013).
\newblock \doi{10.1016/j.physletb.2013.08.030}

\bibitem{DErrico:2011wfa}
L.~D'Errico, P.~Richardson, Eur. Phys. J. C \textbf{72}, 2042 (2012).
\newblock \doi{10.1140/epjc/s10052-012-2042-x}

\bibitem{Azzi:2019yne}
P.~Azzi, et~al., CERN Yellow Rep. Monogr. \textbf{7}, 1 (2019).
\newblock \doi{10.23731/CYRM-2019-007.1}

\bibitem{Jager:2014vna}
B.~J\"ager, F.~Schissler, D.~Zeppenfeld, JHEP \textbf{07}, 125 (2014).
\newblock \doi{10.1007/JHEP07(2014)125}

\bibitem{deFlorian:2016spz}
D.~de~Florian, et~al.,  \textbf{2/2017} (2016).
\newblock \doi{10.23731/CYRM-2017-002}

\bibitem{Sherpa:2019gpd}
E.~Bothmann, et~al., SciPost Phys. \textbf{7}(3), 034 (2019).
\newblock \doi{10.21468/SciPostPhys.7.3.034}

\bibitem{Alioli:2010xd}
S.~Alioli, P.~Nason, C.~Oleari, E.~Re, JHEP \textbf{06}, 043 (2010).
\newblock \doi{10.1007/JHEP06(2010)043}

\bibitem{Hoche:2021mkv}
S.~H\"oche, S.~Mrenna, S.~Payne, C.T. Preuss, P.~Skands,   (2021)

\bibitem{Campanario:2013fsa}
F.~Campanario, T.M. Figy, S.~Plätzer, M.~Sjödahl, Phys. Rev. Lett.
  \textbf{111}(21), 211802 (2013).
\newblock \doi{10.1103/PhysRevLett.111.211802}

\bibitem{Campanario:2018ppz}
F.~Campanario, T.M. Figy, S.~Plätzer, M.~Rauch, P.~Schichtel, M.~Sjödahl,
  Phys. Rev. \textbf{D98}(3), 033003 (2018).
\newblock \doi{10.1103/PhysRevD.98.033003}

\bibitem{Liu:2019tuy}
T.~Liu, K.~Melnikov, A.A. Penin, Phys. Rev. Lett. \textbf{123}(12), 122002
  (2019).
\newblock \doi{10.1103/PhysRevLett.123.122002}

\bibitem{Dreyer:2020urf}
F.A. Dreyer, A.~Karlberg, L.~Tancredi, JHEP \textbf{10}, 131 (2020).
\newblock \doi{10.1007/JHEP10(2020)131}

\bibitem{Berger:2019wnu}
N.~Berger, et~al.,   (2019)

\bibitem{ATLAS:2020bhl}
G.~Aad, et~al., Eur. Phys. J. C \textbf{81}(6), 537 (2021).
\newblock \doi{10.1140/epjc/s10052-021-09192-8}

\bibitem{Buckley:2021gfw}
A.~Buckley, et~al., JHEP \textbf{11}, 108 (2021).
\newblock \doi{10.1007/JHEP11(2021)108}

\bibitem{Bahr:2008pv}
M.~Bahr, et~al., Eur. Phys. J. C \textbf{58}, 639 (2008).
\newblock \doi{10.1140/epjc/s10052-008-0798-9}

\bibitem{Bellm:2015jjp}
J.~Bellm, et~al., Eur. Phys. J. C \textbf{76}(4), 196 (2016).
\newblock \doi{10.1140/epjc/s10052-016-4018-8}

\bibitem{Platzer:2011bc}
S.~Platzer, S.~Gieseke, Eur. Phys. J. C \textbf{72}, 2187 (2012).
\newblock \doi{10.1140/epjc/s10052-012-2187-7}

\bibitem{Bellm:2019zci}
J.~Bellm, et~al., Eur. Phys. J. C \textbf{80}(5), 452 (2020).
\newblock \doi{10.1140/epjc/s10052-020-8011-x}

\bibitem{Campanario:2013nca}
F.~Campanario, T.M. Figy, S.~Plätzer, M.~Sjödahl, PoS \textbf{RADCOR2013},
  042 (2013).
\newblock \doi{10.22323/1.197.0042}

\bibitem{Campanario:2014aia}
F.~Campanario, T.M. Figy, S.~Plätzer, M.~Sjodahl, PoS \textbf{LL2014}, 025
  (2014).
\newblock \doi{10.22323/1.211.0025}

\bibitem{Sjodahl:2014opa}
M.~Sjodahl, Eur. Phys. J. \textbf{C75}(5), 236 (2015).
\newblock \doi{10.1140/epjc/s10052-015-3417-6}

\bibitem{Campanario:2011cs}
F.~Campanario, JHEP \textbf{10}, 070 (2011).
\newblock \doi{10.1007/JHEP10(2011)070}

\bibitem{baglio2011vbfnlo}
J.~Baglio, J.~Bellm, G.~Bozzi, M.~Brieg, F.~Campanario, C.~Englert, B.~Feigl,
  J.~Frank, T.~Figy, F.~Geyer, et~al., arXiv preprint arXiv:1107.4038  (2011)

\bibitem{baglio2014release}
J.~Baglio, J.~Bellm, F.~Campanario, B.~Feigl, J.~Frank, T.~Figy, M.~Kerner,
  L.~Ninh, S.~Palmer, M.~Rauch, et~al., arXiv preprint arXiv:1404.3940  (2014)

\bibitem{Arnold:2008rz}
K.~Arnold, et~al., Comput. Phys. Commun. \textbf{180}, 1661 (2009).
\newblock \doi{10.1016/j.cpc.2009.03.006}

\bibitem{Alioli:2013nda}
S.~Alioli, et~al., Comput. Phys. Commun. \textbf{185}, 560 (2014).
\newblock \doi{10.1016/j.cpc.2013.10.020}

\bibitem{Binoth:2010xt}
T.~Binoth, et~al., Comput. Phys. Commun. \textbf{181}, 1612 (2010).
\newblock \doi{10.1016/j.cpc.2010.05.016}

\bibitem{Cacciari:2005hq}
M.~Cacciari, G.P. Salam, Phys. Lett. \textbf{B641}, 57 (2006).
\newblock \doi{10.1016/j.physletb.2006.08.037}

\bibitem{Buckley:2014ana}
A.~Buckley, J.~Ferrando, S.~Lloyd, K.~Nordstr\"om, B.~Page, M.~R\"ufenacht,
  M.~Sch\"onherr, G.~Watt, Eur. Phys. J. C \textbf{75}, 132 (2015).
\newblock \doi{10.1140/epjc/s10052-015-3318-8}

\bibitem{Butterworth_2016}
J.~Butterworth, S.~Carrazza, A.~Cooper-Sarkar, A.D. Roeck, J.~Feltesse,
  S.~Forte, J.~Gao, S.~Glazov, J.~Huston, Z.~Kassabov, et~al., Journal of
  Physics G: Nuclear and Particle Physics \textbf{43}(2), 023001 (2016).
\newblock \doi{10.1088/0954-3899/43/2/023001}.
\newblock \urlprefix\url{http://dx.doi.org/10.1088/0954-3899/43/2/023001}

\bibitem{Cacciari:2011ma}
M.~Cacciari, G.P. Salam, G.~Soyez, Eur. Phys. J. \textbf{C72}, 1896 (2012).
\newblock \doi{10.1140/epjc/s10052-012-1896-2}

\bibitem{Buckley_2013}
A.~Buckley, J.~Butterworth, D.~Grellscheid, H.~Hoeth, L.~Lönnblad, J.~Monk,
  H.~Schulz, F.~Siegert, Computer Physics Communications \textbf{184}(12),
  2803–2819 (2013).
\newblock \doi{10.1016/j.cpc.2013.05.021}.
\newblock \urlprefix\url{http://dx.doi.org/10.1016/j.cpc.2013.05.021}

\bibitem{HJets:merging}
T.~Chen, T.~Figy.
\newblock {Herwig 7 Simulations using VBFNLO and HJETS Matrix Elements} (2021).
\newblock \urlprefix\url{https://github.com/drtmfigy/herwig-hjets-merging}

\bibitem{Catani:1993hr}
S.~Catani, Y.L. Dokshitzer, M.H. Seymour, B.R. Webber, Nucl. Phys. B
  \textbf{406}, 187 (1993).
\newblock \doi{10.1016/0550-3213(93)90166-M}

\bibitem{Ellis:1993tq}
S.D. Ellis, D.E. Soper, Phys. Rev. D \textbf{48}, 3160 (1993).
\newblock \doi{10.1103/PhysRevD.48.3160}

\bibitem{ATLAS:2013nef}
G.~Aad, et~al., Eur. Phys. J. C \textbf{73}(5), 2432 (2013).
\newblock \doi{10.1140/epjc/s10052-013-2432-8}

\bibitem{Bellm:2016rhh}
J.~Bellm, G.~Nail, S.~Pl\"atzer, P.~Schichtel, A.~Si\'odmok, Eur. Phys. J. C
  \textbf{76}(12), 665 (2016).
\newblock \doi{10.1140/epjc/s10052-016-4506-x}

\bibitem{Platzer:2009jq}
S.~Platzer, S.~Gieseke, JHEP \textbf{01}, 024 (2011).
\newblock \doi{10.1007/JHEP01(2011)024}

\bibitem{Gieseke:2003rz}
S.~Gieseke, P.~Stephens, B.~Webber, JHEP \textbf{12}, 045 (2003).
\newblock \doi{10.1088/1126-6708/2003/12/045}

\bibitem{Frixione:2002ik}
S.~Frixione, B.R. Webber, JHEP \textbf{06}, 029 (2002).
\newblock \doi{10.1088/1126-6708/2002/06/029}

\bibitem{Nason:2004rx}
P.~Nason, JHEP \textbf{11}, 040 (2004).
\newblock \doi{10.1088/1126-6708/2004/11/040}

\bibitem{Lonnblad:2011xx}
L.~Lonnblad, S.~Prestel, JHEP \textbf{03}, 019 (2012).
\newblock \doi{10.1007/JHEP03(2012)019}

\bibitem{Lonnblad:2012ix}
L.~L\"onnblad, S.~Prestel, JHEP \textbf{03}, 166 (2013).
\newblock \doi{10.1007/JHEP03(2013)166}

\bibitem{Lonnblad:2012ng}
L.~Lonnblad, S.~Prestel, JHEP \textbf{02}, 094 (2013).
\newblock \doi{10.1007/JHEP02(2013)094}

\bibitem{Bellm:2017ktr}
J.~Bellm, S.~Gieseke, S.~Pl\"atzer, Eur. Phys. J. C \textbf{78}(3), 244 (2018).
\newblock \doi{10.1140/epjc/s10052-018-5723-2}

\bibitem{Catani:1990rr}
S.~Catani, B.~Webber, G.~Marchesini, Nucl. Phys. B \textbf{349}, 635 (1991).
\newblock \doi{10.1016/0550-3213(91)90390-J}

\bibitem{Hoche:2019flt}
S.~H\"oche, S.~Prestel, H.~Schulz, Phys. Rev. D \textbf{100}(1), 014024 (2019).
\newblock \doi{10.1103/PhysRevD.100.014024}

\bibitem{Hamilton:2012np}
K.~Hamilton, P.~Nason, G.~Zanderighi, JHEP \textbf{10}, 155 (2012).
\newblock \doi{10.1007/JHEP10(2012)155}

\bibitem{Ferrera:2013yga}
G.~Ferrera, M.~Grazzini, F.~Tramontano, JHEP \textbf{04}, 039 (2014).
\newblock \doi{10.1007/JHEP04(2014)039}

\bibitem{Ferrera:2014lca}
G.~Ferrera, M.~Grazzini, F.~Tramontano, Phys. Lett. B \textbf{740}, 51 (2015).
\newblock \doi{10.1016/j.physletb.2014.11.040}

\bibitem{Ferrera:2014yda}
G.~Ferrera, PoS \textbf{LL2014}, 064 (2014).
\newblock \doi{10.22323/1.211.0064}

\end{thebibliography}

\end{document}